\documentclass[aps,prb,reprint,amssymb,amsmath,superscriptaddress]{revtex4-2}
\usepackage{graphicx}
\usepackage{epstopdf}

\begin{document}

\title{Quantum oscillations of transport coefficients and capacitance: an unexpected manifestation of the spin-Hall effect}

\author{G. M.~Minkov}

\affiliation{School of Natural Sciences and Mathematics, Ural Federal University,
620002 Ekaterinburg, Russia}

\affiliation{M.~N.~Mikheev Institute of Metal Physics of Ural Branch of
Russian Academy of Sciences, 620137 Ekaterinburg, Russia}

\author{O.\,E.~Rut}
\affiliation{School of Natural Sciences and Mathematics, Ural Federal University,
620002 Ekaterinburg, Russia}

\author{A.\,A.~Sherstobitov}

\affiliation{School of Natural Sciences and Mathematics, Ural Federal University,
620002 Ekaterinburg, Russia}

\affiliation{M.~N.~Mikheev Institute of Metal Physics of Ural Branch of
Russian Academy of Sciences, 620137 Ekaterinburg, Russia}

\author{S.\,A.~Dvoretski}

\affiliation{Institute of Semiconductor Physics RAS, 630090
Novosibirsk, Russia}

\author{N.\,N.~Mikhailov}

\affiliation{Institute of Semiconductor Physics RAS, 630090
Novosibirsk, Russia}
\affiliation{Novosibirsk State University, Novosibirsk 630090, Russia}

\author{A.~V.~Germanenko}

\affiliation{School of Natural Sciences and Mathematics, Ural Federal University,
620002 Ekaterinburg, Russia}

\date{\today}

\begin{abstract}
The results of systematic experimental studies of quantum oscillations of  resistivity,  Hall coefficient and capacitance in GaAs and In$_x$Ga$_{1-x}$As quantum wells (QWs) with a simple electron spectrum and HgTe QWs with a complicated non-parabolic spectrum and strong spin-orbit interaction are reported. A striking result is the ratio of the amplitudes of the resistance and Hall coefficient oscillations. In GaAs QW with a simple spectrum characterized by negligibly small Zeeman and spin-orbit splitting, the ratio of amplitudes is close to that predicted theoretically. In HgTe QWs, this ratio is very different and behaves differently in QWs with normal and inverted electron spectra. In HgTe WQs with a normal spectrum, it tends to a theoretical value with an increase of the filling factor ($N$), while for HgTe QWs with an inverted spectrum, it differs significantly from the theoretical one for all available $N$. It is assumed that such a difference in the ratio of amplitudes in GaAs and HgTe QWs is due not to the peculiarities of the energy spectrum of HgTe, but to the peculiarities of electron scattering due to spin-orbit interaction with the potential of the scatterers. This assumption is justified by analysis of experimental results obtained for a heterostructure with a In$_{0.2}$Ga$_{0.8}$As QW, which spectrum is very close to the GaAs QW spectrum, but characterizes by much stronger spin-orbit splitting value. It has been found that the positions of the resistance and capacitance oscillations, the difference between the phases of the resistance and Hall coefficient oscillations and its $N$ dependence are close to those observed in GaAs QW. At the same time the ratio of the amplitude of the resistance oscillations to the Hall coefficient oscillations and its $N$ dependence differs very strongly and they are close to that observed in HgTe quantum wells.
\end{abstract}


\maketitle
\section{Introduction}
\label{sec:intr}

The study of Shubnikov-de Haas (SdH) oscillations is one of the most common and powerful methods for studying the energy spectrum, carrier scattering mechanisms in semiconductors, metals and semimetals. The possibilities of this method are significantly expanded when studying two-dimensional (2D) systems, since the gate electrode allows us to change the carrier density $(n)$, and hence the Fermi energy $(\varepsilon_F)$ in a wide range.

The SdH oscillations result from the quantization of the energy spectrum by the magnetic field $(B)$ that leads to oscillations of the density of states $(\nu)$ at a fixed energy as a function of the magnetic field, and/or to oscillations of the density of states with energy at a fixed magnetic field. Oscillations in the density of states lead to oscillations in the momentum relaxation time $\tau_B$.

Theoretically, the SdH oscillations have been considered in many papers \cite{LifKos55,Ando74,Ando75,Isihara86,Novokshonov13}. As a rule, the components of the conductivity tensor $\sigma$ were calculated in them. Established expressions $\sigma_{xx}$ and $\sigma_{xy}$ \cite{Dmitriev12} are:
\begin{eqnarray}
\sigma_{xx}&=&e^2v_F^2\int d\varepsilon\left(-\frac{\partial f_\varepsilon^T}{\partial\varepsilon}\right)\frac{\nu(\varepsilon)\tau_B(\varepsilon)}{1+\omega_c^2\tau_B^2} \label{eq10} \\
\sigma_{xy}&=&-\frac{en}{B}+\frac{e^2v_F^2}{\omega_c}\int d\varepsilon\left(-\frac{\partial f_\varepsilon^T}{\partial\varepsilon}\right)\frac{\nu(\varepsilon)}{1+\omega_c^2\tau_B^2}, \label{eq20}
\end{eqnarray}
where $e$ is the elementary charge, $f_\varepsilon^T$
is the thermal distribution function, $\omega_c=eB/m_e$, $\nu(\varepsilon)=\nu_0 (1+ \Delta\nu(\varepsilon)/\nu_0$) is the density of states with  $\nu_0$ as the density of states at $B=0$, and $\Delta\nu(\varepsilon)$ as the oscillating part of the density of states, $v_F$ is the carrier velocity at the Fermi level.
Expressions (\ref{eq10}) and (\ref{eq20}) can be written in a simpler form if $\varepsilon_F\gg k_BT$:
\begin{eqnarray}
\sigma_{xx}(B,\varepsilon_F)&=&\frac{e^2v_F^2\tau_B(B,\varepsilon_F) \nu(B,\varepsilon_F)}{1+\omega_c^2(B)\tau_B^2(B,\varepsilon_F)} \label{eq30} \\
\sigma_{xy}&=&-\frac{en}{B}+\frac{e^2v_F^2}{\omega_c(B)} \frac{ \nu(B,\varepsilon_F)}{1+\omega_c^2(B)\tau_B^2(B,\varepsilon_F)}. \label{eq40}
\end{eqnarray}

The result is that the theory for the first subband of spatial quantization   can be formulated solely in terms of the oscillating $\Delta\nu$ and the transport scattering rate $1/\tau(B)$ which is renormalized by Landau quantization. The renormalization of the density of states $\nu(B,\varepsilon_F)$ and the relaxation time $\tau_B (B,\varepsilon_F)$ in a magnetic field are related as
$1/\tau_B(B,\varepsilon_F)=1/\tau\times \nu(B,\varepsilon_F)/\nu_0$
\cite{Dmitriev12}, where $\tau$ is the transport relaxation time in zero magnetic field. Considering this, we get for $\sigma_{xx}$
\begin{equation}
\label{eq50}
\sigma_{xx}(B,\varepsilon_F)=\frac{e^2v_F^2\tau \nu_0}{1+\omega_c^2(B)\tau_B^2(B,\varepsilon_F)}
\end{equation}
As a rule, the components of the resistance tensor $\rho$ are measured experimentally, which are related to the $\sigma$ components  by simple relations $\rho_{xx}=\sigma_{xx}⁄(\sigma_{xx}^2+\sigma_{xy}^2 ) $   and  $\rho_{xx}=\sigma_{xy}⁄(\sigma_{xx}^2+\sigma_{xy}^2  )$. Expanding  Eqs.~(\ref{eq40}) and (\ref{eq50}) in terms of the smallness of the amplitude of oscillations of the density of states $\Delta\nu/\nu_0$, one obtains
\begin{eqnarray}
\rho_{xx}&=&\rho_0\left(1+2\frac{\Delta\nu}{\nu_0}\right), \label{eq60} \\
\rho_{xy}&=&\frac{B}{en}\left(1-\frac{1}{\mu^2 B^2}\frac{\Delta\nu}{\nu_0}\right). \label{eq70}
\end{eqnarray}
Here $\mu=e\tau/m$ is the mobility, $\rho_0=1/(en\mu)$. Oscillating parts of $\rho_{xx}$ and $R_H=-\rho_{xy}/B$, defined as $\delta\rho_{xx}=\rho_{xx}/\rho_0 - 1$ and $\delta R_H=R_H/R_0-1$, where $R_0=-1/en$, have the form
\begin{eqnarray}
\delta\rho_{xx}&=&2\frac{\Delta\nu}{\nu_0}\mathcal{F}\left(\frac{2\pi^2k_BT}{\hbar\omega_c}\right), \label{eq80} \\
\delta R_H&=&-\frac{1}{(\mu B)^2}\frac{\Delta\nu}{\nu_0}\mathcal{F}\left(\frac{2\pi^2k_BT}{\hbar\omega_c}\right), \label{eq90}
\end{eqnarray}
where the factor $\mathcal{F}(x)=x/\sinh(x)$ describes the thermal averaging of the oscillations,
\begin{equation}
\label{eq92}
\frac{\Delta\nu}{\nu_0}=-2\exp\left(-\frac{\pi}{\omega_c\tau_q}\right)\cos\left( \frac{2\pi \varepsilon_F}{\hbar\omega_c}\right)
\end{equation}
for the case when the Fermi level is kept constant in the magnetic field, and
\begin{equation}
\label{eq93}
\frac{\Delta\nu}{\nu_0}=-2\exp\left(-\frac{\pi}{\omega_c\tau_q}\right)\cos\left( \frac{2\pi^2 \hbar\, n }{eB}\right),
\end{equation}
when the electron density is constant, therewith the Landau levels are  twofold degenerate. Here, $\tau_q$ is the quantum relaxation time. Since $n=\text{const}_B$ in the structures under study (see Section~\ref{sec:RoandC}) we will use Eq.~(\ref{eq93}) so that
\begin{equation}
\label{eq97}
\delta\rho_{xx}=-4\exp\left(-\frac{\pi}{\omega_c\tau_q}\right)\cos\left( \frac{2\pi^2 \hbar\, n }{eB}\right)\mathcal{F}\left(\frac{2\pi^2k_BT}{\hbar\omega_c}\right).
\end{equation}
It follows From Eqs.~(\ref{eq80}) and (\ref{eq90}) that the oscillations of $\rho_{xx}$ and $R_H$ should be in antiphase, and the ratio of the oscillation amplitudes $\rho_{xx}$ and $R_H$ (denoted as $A_\rho$ and $A_H$, respectively) should be equal to
\begin{equation}
\label{eq100}
\frac{A_\rho}{A_H}=-2\mu^2 B^2.
\end{equation}

The approximations made in obtaining the equations from (\ref{eq60}) to (\ref{eq100}) are discussed in detail in Ref.~\cite{Dmitriev12}. They are as follows:
\begin{enumerate}
\item
The equations  are valid in low magnetic fields as long as $\delta\rho_{xx}\ll 1 $, i.e., before the onset of the quantum Hall effect (QHE) regime;
\item
There is neither Zeeman nor spin-orbital (SO) splitting of Landau levels in the actual magnetic field range;
\item
It is assumed that there is only a homogeneous broadening of the Landau levels, which leads to a Lorentzian shape of the density of states with a width $\Delta_L$. That is, large-scale fluctuations, which lead to a Gaussian shape of the density of states with a width $\Delta_G$, are neglected.
\end{enumerate}

Note that the mobility $\mu$ in Eq.~(\ref{eq100}) includes the transport relaxation time, whereas the width of the Landau levels is determined by the quantum relaxation time. These times coincide only in the case of a short-range scattering potential when the Gaussian broadening is absent.

Eqs. (\ref{eq80}) and (\ref{eq90}) are very simple, but a lot follows from them, namely:
\begin{enumerate}
\item Oscillations are periodic in the reciprocal magnetic field;
\item The period of oscillations in the reciprocal magnetic field is determined by the carrier density and the degeneracy of the states;
\item  Comparison of this density with the Hall density $n_H=1⁄(e|R_H| )$ makes it possible to determine the degeneracy  of the Landau levels;
\item  Predicted dependence of the oscillation amplitude on temperature and magnetic field makes it possible to determine the effective mass of carriers the broadening of the Landau levels;
\item  The positions of the oscillations of $\rho_{xx}(B)$ should coincide with the positions of the oscillations of the density of states.
\item The $\delta\rho_{xx}$ and $\delta R_H$  oscillations should be antiphase;
\item  Ratio of oscillation amplitudes of $\delta\rho_{xx}$ and $\delta R_H$ and their dependences on the magnetic field should satisfy Eq.~(\ref{eq100});
\item Extrapolating the positions of the oscillation minima plotted in the $\rho_{xx}$ vs $1/B$ coordinates  to $B=0$ allows one to determine the Berry phase\footnote{As seen from Eq.~(\ref{eq80}), the magnetic field of the $N$th $\rho_{xx}$ minimum  ($B_N$) corresponds to the minimum of the density of states at the Fermi level, when the integer number of Landau levels are occupied, i.e., the Fermi level is located between the two Landau levels. Note,  the values of $B_N$ are different for the two regimes: $E_F=\text{const}_B$ and $n=\text{const}_B$. In the first case, taking into account the fact that the energy of the Landau levels is $E_N=\hbar\omega_c(N+1/2+\phi)$, where $\phi$ is the Berry phase, the fields of $\rho_{xx}$ minima  are determined by the condition $\varepsilon_F=E_N+\hbar\omega_c/2= e\hbar B_N/m\times (N+1/2+1/2+\phi)$. Thus we get $B_N=\varepsilon_F/[e\hbar/m (N+1+\phi)]$, i.e., an extrapolation of the $N$ versus $1/B_N$ experimental plot allows us to obtain the Barry phase. In the second case, $n=\text{const}_B$, the condition for filling an integer number of Landau levels is simpler: the number of states at twofold degenerate Landau levels in the field $B$ is $eB/(\pi \hbar)$, so $B_N=\pi \hbar n/(eN)$, This shows that at $n=\text{const}_B$ $B_N$ does not include the Berry phase \cite{Kuntsevich18}.}.
\end{enumerate}

Experimentally, the SdH oscillations have been studied in a very wide range of 2D systems. The discrepancy between the experimental and predicted value of the phase difference of the $\rho_{xx}$ and $R_{H}$ oscillations, found in many papers, poses a simple question not answered up to now: the phase of which oscillations, $\rho_{xx}$  or $R_H$, differs from that predicted theoretically? There is no experimental answer to it. As mentioned above the positions of the oscillations of $\rho_{xx}(B)$ should coincide with the positions of the oscillations of the density of states $\nu$.  On the other hand, the oscillations of $\nu$ directly determine the oscillations of capacitance ($C$) between the two-dimensional electron gas and the gate electrode. Hence a comparison of the oscillations  of $\rho_{xx}(B)$ with those of $C(B)$ can give an unambiguous answer to this question.

The relationship between the  amplitudes of $\rho_{xx}$ and $R_H$ oscillations and their magnetic field dependences, as far as we know, was studied only in \cite{Coleridge89} for electrons in GaAs quantum well (QW) with the simplest spectrum, small Zeeman and SO splitting.

In this paper we present the results of the experimental investigations of  the oscillations of the resistivity, Hall coefficient, and capacitance between the gate electrode and quantum well in GaAs and In$_x$Ga$_{1-x}$As QWs with a simple spectrum of the conduction band and in HgTe QWs with a complicated non-parabolic spectrum and strong SO interaction.

The paper is organized as follows. The samples and experimental conditions are described in the next section. The experimental results and their analysis concerning the phase of the oscillations of $\rho_{xx}$,  $C$, and $R_H$  for the GaAs and HgTe QWs are presented in Secs.~\ref{sec:RoandC} and \ref{sec:phase}. Section~\ref{sec:ampl} is devoted to the analysis of the magnetic field dependences of the ratio of the amplitudes of the quantum oscillations of $\rho_{xx}$ and $R_H$. The discussion is given  in Section~\ref{sec:disc}. Also in this section, the results obtained for the In$_{x}$Ga$_{1-x}$As QW are presented to elucidate the role of the physical mechanisms responsible for the effects considered in the previous sections.  The conclusions are in Sec.~\ref{sec:concl}.

\section{Experimental}
\label{sec:exp}

Our samples with the HgTe quantum wells  were realized on the basis of HgTe/Hg$_{1-x}$Cd$_{x}$Te ($x=0.39-0.6$) heterostructures grown by the
molecular beam epitaxy on a GaAs substrate \cite{Mikhailov06}.

The quantum well GaAs/In$_{0.2}$Ga$_{0.8}$As/GaAs heterostructure
were grown by metal-organic vapor phase epitaxy on semi-insulator
GaAs substrate. The heterostructure consists of 0.5 mkm-thick undoped GaAs epilayer, a Sn $\delta$-layer, a $12$ nm spacer of undoped GaAs, a 8 nm
In$_{0.2}$Ga$_{0.8}$As well, a $12$ nm spacer of undoped GaAs, a Sn
$\delta$-layer, and a 300 nm cap layer of undoped GaAs.

The Al$_x$Ga$_{1-x}$As/GaAs/Al$_x$Ga$_{1-x}$As
heterostructure was grown by molecular beam epitaxy. It consists of
$250$~nm-thick undoped GaAs buffer layer grown on semiinsulator
GaAs, a $50$~nm Al$_{0.3}$Ga$_{0.7}$As barrier, Si $\delta$-layer, a
$6$~nm spacer of undoped Al$_{0.3}$Ga$_{0.7}$As, a $8$~nm GaAs well,
a $6$~nm spacer of undoped Al$_{0.3}$Ga$_{0.7}$As, a Si
$\delta$-layer, a $50$~nm Al$_{0.3}$Ga$_{0.7}$As barrier, and
$200$~nm cap layer of undoped GaAs.

The samples were mesa etched into standard Hall bars of $0.5$-mm width with the distance between the potential probes of $0.5$~mm. To change and control the carrier density in the quantum well, the field-effect transistors were fabricated with parylene as an insulator and aluminium as a gate electrode. For each heterostructure, several samples were fabricated and studied. All measurements were carried out using the \emph{dc} technique in the linear response regime at $T=(1.3\ldots 10.0)$~K within the magnetic field range up to $2.5$~T.

The parameters of the structures under study are presented in the Table~\ref{tab1}.

\begin{table*}
\caption{The parameters of  heterostructures under study}
\label{tab1}
\begin{ruledtabular}
\begin{tabular}{cccccccc}
\# & QW &  $d$ (nm) & type & $n$(cm$^{-2}$)$^\text{a}$& $\mu$ (m$^2$/Vs)$^\text{a}$\\
\colrule
 T1520 & GaAs & $8.0$   &$n$  & $1.56\times10^{12}$ &   $1.9$ \\
  170928 & HgTe & $5.1$   &$n$  &  $4.32\times10^{11}$ &  $2.9$  \\
 H1023 & HgTe & $6.5$   &$n$  &  $1.60\times10^{11}$ &  $9.5$  \\
 HT71 & HgTe & $9.5$   & $n$ &    $0.93\times10^{11}$ & $6.3$ \\
 180824 & HgTe & $32$   & $p$ &    $-$ & $-$ \\
 Z76 & In$_{0.2}$Ga$_{0.8}$As & $8.0$   & $n$ &    $6.25\times10^{11}$ & $2.5$ \\
\end{tabular}
\end{ruledtabular}
\footnotetext[1]{For $V_g=0$~V}
\end{table*}

\section{Phase of  oscillations of $\rho_{xx} (B)$ and $C(B)$}
\label{sec:RoandC}

Let us try to answer the simplest question formulated in the introduction: does the phase of oscillations of $\rho_{xx} (B)$ coincide with the phase  of oscillations of the density of states?

\begin{figure}
\includegraphics[width=1.0\linewidth,clip=true]{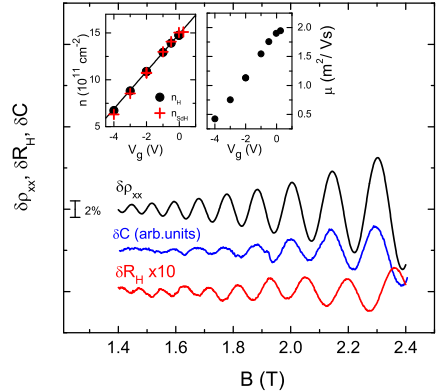}
\caption{(color online) Magnetic field dependences  of $\delta\rho_{xx}$,  $\delta R_H$ and of the oscillating part of the capacitance $\delta C$ at $V_g=0$, $T=3.9$~K. The left inset shows the gate voltage dependences of the Hall density $n=1/e|R_H(1\text{~T})|$ and the electron density determined from the period of SdH oscillations under the assumption of twofold degeneracy of the Landau levels. The line is the dependence $n=(14.9+2.0\,V_g)\times 10^{11}$~cm$^{-2}$. The right inset is the gate voltage dependences  of the mobility determined as  $\mu=|R_H(1\text{~T})|/\rho_{xx}(1\text{ T})$. The structure T1520 with GaAs QW. }
\label{f1}
\end{figure}

First we consider the results of oscillation measurements for a structure T1520 with GaAs QW  with the simplest spectrum, in which the Zeeman and SO splitting is negligibly small in the used range of magnetic fields, temperatures, and electron density. For a general characterization of the structure, let us inspect the insets of Fig.~\ref{f1}.

The left inset in Fig.~\ref{f1} shows that the electron densities found from the Hall effect and from the oscillation period under the assumption of a twofold degeneracy of Landau levels coincide. The electron density changes linearly with $V_g$ with a rate of $2.0\times 10^{11}$~cm$^{-2}$V$^{-1}$. This value, within the experimental error, coincides with $dn/dV_g$ determined from the capacitance value. These results show that there are no additional states in the barriers near the Fermi level in this range of electron density, which also implies that the electron density remains constant when the magnetic field changes.

Fig.~\ref{f1} shows the oscillating parts of $\rho_{xx}(B)$, $R_H(B)$ (obtained as  $\delta\rho_{xx}=\rho_{xx}/\rho_{xx}^{mon}-1$, $\delta R_H=R_H/R_H^{mon}-1$, where, $\rho_{xx}^{mon}$ and $R_H^{mon}$ are the monotonic parts of the magnetic field dependences of $\rho_{xx}$ and  $R_H$, respectively) \footnote{in the equations given in Section~\ref{sec:intr}, only the short-range potential scattering  is taken into account, so there are no monotonous parts of the dependences $\rho_{xx}(B)$, $R_H(B)$. In real structures, there are various mechanisms that can lead to their appearance, for instance, the  scattering by potential fluctuations, the electron-electron interaction, etc.}, and  oscillating part of $C(B)$ measured on the same structure under the same conditions for $n\simeq 1.5\times 10^{12}$~cm$^{-2}$.  As seen the phase of $\delta\rho_{xx}(B)$ differs from the phase of $\delta R_H(B)$, but this difference is not equal to $\pi$. The main thing clearly seen from this figure is that the positions of the oscillations of $\delta\rho_{xx}(B)$ and $\delta C(B)$ coincide. Since the $C(B)$ oscillations are directly determined by the oscillations of the density of states at the Fermi level it follows from this that the oscillations $\rho_{xx}(B)$ are also determined by the oscillations of the density of states at the Fermi level only. The same ratio of the positions of the $C(B)$ and $\rho_{xx}(B)$ oscillations is also observed at lower densities, and, despite the fact that the amplitude of the oscillations becomes much smaller, the mutual positions of the extrema of $C(B)$ and $\rho_{xx}(B)$ remain the same.

Let us now consider the results obtained for structures with HgTe QWs which spectrum changes very strongly with increasing QW width  from the normal at $d<6.3$~nm to inverted at $d>6.3$~nm and semimetallic at $d\gtrsim 15$~nm \footnote{As shown in numerous papers (see, e.g., \cite{Gerchikov90,Zhang01,Novik05,Bernevig06,ZholudevPhD,Ren2016} and references therein) different types of spectrum are realized depending on the QW width.   When  QW is narrow, $d < d_c$, the ordering of energy subbands of spatial quantization is analogous to that in conventional semiconductors; the highest valence subband at $k = 0$ is
formed from the heavy hole $\Gamma_8$ states, while the lowest conduction
subband is formed both from the $\Gamma_6$ states and light hole $\Gamma_8$
states. For a thicker HgTe layer, $d > d_c$, the quantum well is in
the inverted regime; the lowest conduction subband is formed from
the heavy hole $\Gamma_8$ states \cite{Dyak82e}, whereas the subband formed
from the $\Gamma_6$ states and light hole $\Gamma_8$ states sinks into the
valence band. In the inverted regime,  the spectrum becomes semimetallic at $d\gtrsim 15$~nm due to overlapping of the valence and conduction bands.}

\begin{figure}
\includegraphics[width=1.0\linewidth,clip=true]{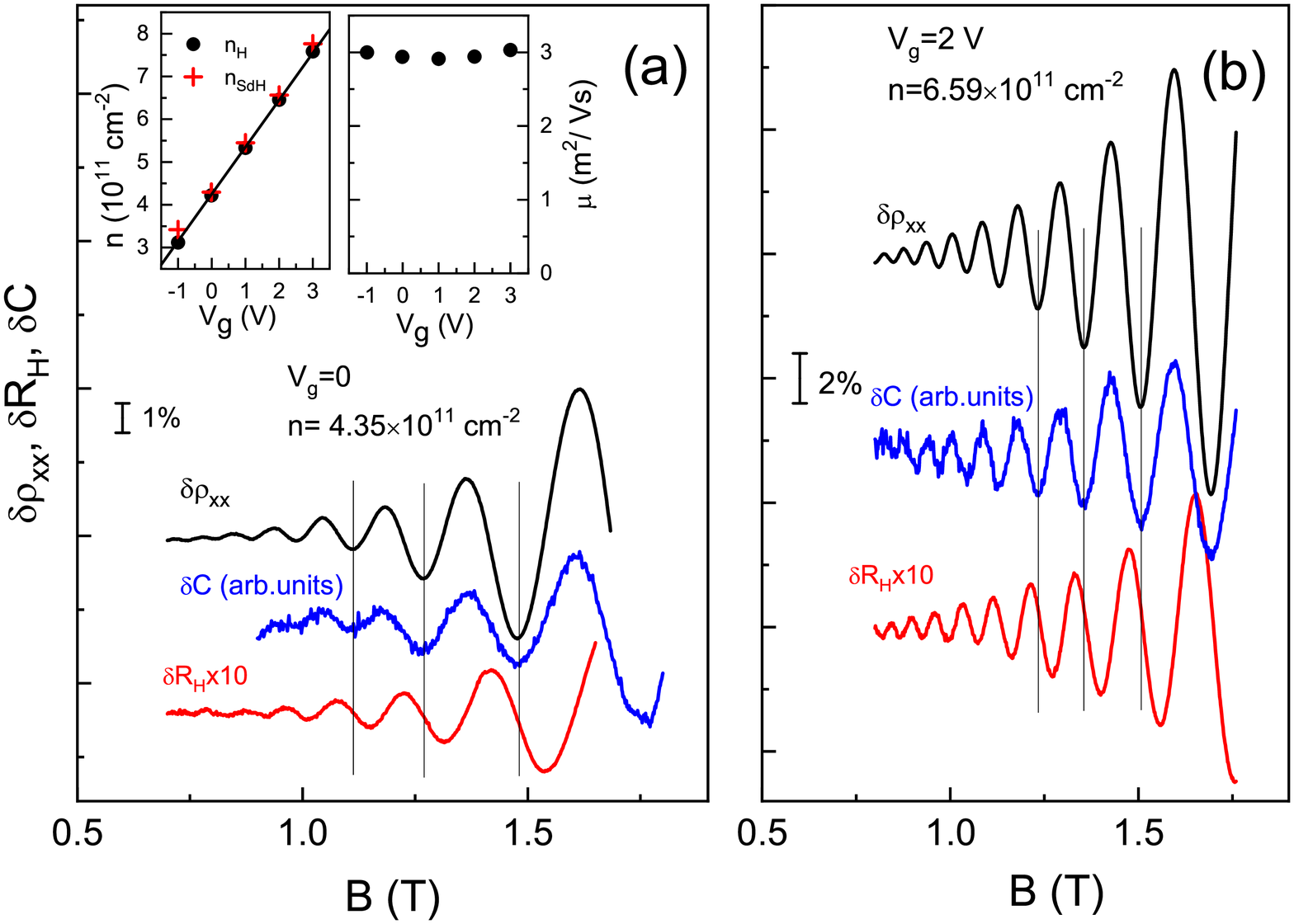}
\caption{(color online) Magnetic field dependences  of $\delta\rho_{xx}$,  $\delta R_H$, and of the oscillating part of the capacitance $\delta C$   for the structure 170928 with HgTe QW  with $d=5.1$~nm for electron density $4.35\times 10^{11}$~cm$^{-2}$ (a) and  $6.59\times 10^{11}$~cm$^{-2}$ (b).
The left inset in (a) shows the gate voltage dependences of the Hall density and the electron density determined from the period of SdH oscillations under the assumption of twofold degeneracy of the Landau levels.  The line  is the dependence $n=(4.25+1.1\,V_g)\times 10^{11}$~cm$^{-2}$. The right inset is the gate voltage dependences  of the mobility.  $T=4.5$~K.}
\label{f2}
\end{figure}

We start by considering the results for  the structure 170928 with a HgTe QW  of $5.1$~nm wide with a normal spectrum (Fig.~\ref{f2}). As in GaAs QWs, the electron density found from the Hall effect and SdH oscillations coincide, varying linearly with $V_g$ [left inset in Fig.~\ref{f2}(a)]. It is seen that the phase of $\delta\rho_{xx}(B)$ differs from the phase of $\delta R_H(B)$, but this difference is not equal to $\pi$. Like GaAs QW, the positions of the oscillations of $\delta\rho_{xx}(B)$ and $\delta C(B)$ coincide for both electron densities.

\begin{figure}
\includegraphics[width=0.7\linewidth,clip=true]{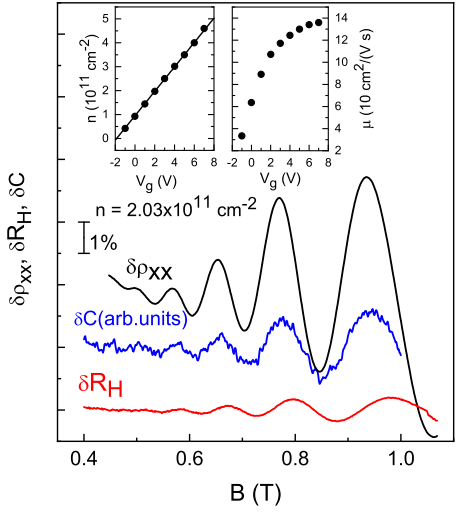}
\caption{(color online) Magnetic field  dependences of $\delta\rho_{xx}$,  $\delta R_H$, and $\delta C$ for the structure HT71 with HgTe QW,  $d=9.5$~nm for $n=2.03\times 10^{11}$~cm$^{-2}$.
The left inset in (a) shows the gate voltage dependences of the Hall density and the electron density determined from the period of SdH oscillations under the assumption of twofold degeneracy of the Landau levels.  The line in the left inset is the dependence $n=(0.95+0.51\,V_g)\times 10^{11}$~cm$^{-2}$. The right inset is the gate voltage dependences  of the mobility.}
\label{f3}
\end{figure}

The same phase coincidence of the $\rho_{xx}(B)$ and $C(B)$ oscillations is observed for the structure HT71 with HgTe QW  with the inverted spectrum, $d=9.5$~nm (Fig.~\ref{f3}).

\begin{figure}
\includegraphics[width=1.0\linewidth,clip=true]{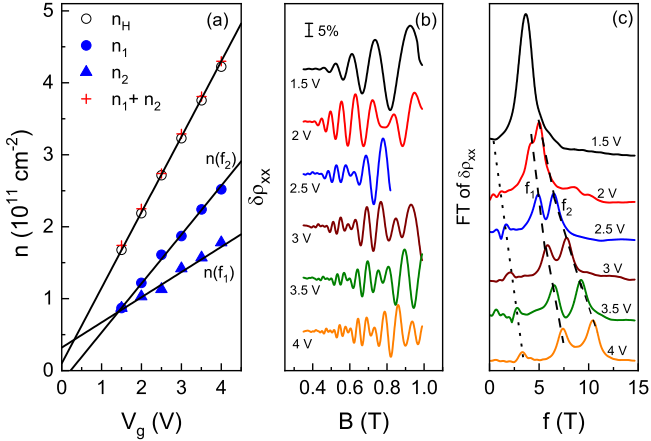}
\caption{(color online) (a) The gate voltage dependence of the electron density for the  structure  180824  with $d=32$~nm: $n_H$ is the Hall density $n_H= 1 / eR_H (0.2\text{ T})$; $n(f_1)$ and $n(f_2)$ are the densities corresponding to the frequencies $f_1$ and $f_2$ [see the panel (c)] considering them come from single-spin Landau levels; lines are the result of the liner fit.  $T=4.2$~K. (b) Oscillations of $\delta\rho_{xx}$ for some $V_g$ values; (c) The Fourier spectra of the oscillations shown in Fig.~\ref{f4}(b) in the $B$ range from $0.2$ to $0.6$~T. The upper limit of B=0.6 T is determined by the onset of QHE. The dotted line in (c) follow the peak corresponding to the magneto-intersubband oscillations \cite{Minkov20-2} not discussed in this paper. }
\label{f4}
\end{figure}

Above we considered the results for HgTe QWs, in which SO interaction  did not lead to spectrum splitting in the actual gate-voltage range. In what follows  we present the results obtained for the structure 180824 with a $32$ nm wide HgTe QW, in which, as the gate voltage increases, a large splitting of the conduction band spectrum occurs due to the spin–orbit interaction of the Rashba type. As  seen from Fig.~\ref{f4}, the Fourier spectrum of $\rho_{xx}$ oscillations  splits into two components of the frequencies $f_1$ and $f_2$ at $V_g>1.5$~V, when the  electron density is higher than $1.7\times 10^{11}$~cm$^{-2}$, and the splitting value increases with increasing $V_g$. In this case, the sum of densities  found from two harmonics, $n(f_1)$, $n(f_2)$, assuming that they are single-spin, is equal to the total density determined from the Hall effect [Fig.~\ref{f4}(a)]. Thus, at $n>1.7\times 10^{11}$~cm$^{-2}$, two single-spin branches are formed in the spectrum of the conduction band.

\begin{figure}
\includegraphics[width=0.9\linewidth,clip=true]{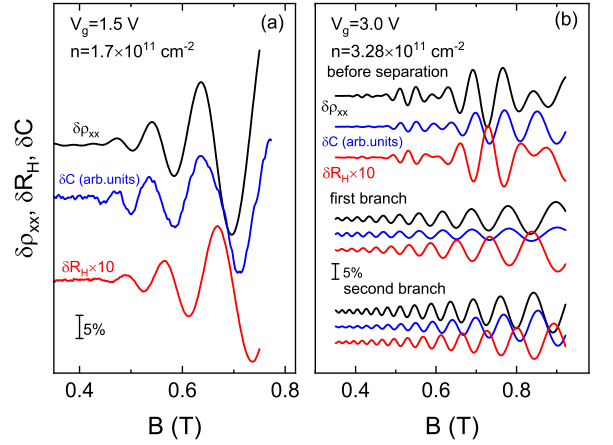}
\caption{(color online) Oscillating parts of $\rho_{xx}$,  $R_H$, and $C$ plotted against the magnetic field for the  structure 180824 with HgTe QW, $d=32$~nm for $V_g=1.5$~V (a) and  $V_g=3.0$~V (b). In Fig.~\ref{f5}(b), the upper group of the curves is the experimental dependences, the middle and lower groups are the results of the inverse Fourier transformation, which correspond to the two branches of the spectrum split by SO interaction.  $T= 4.2$~K. }
\label{f5}
\end{figure}

Let us now consider the results obtained at $V_g=1.5$~V, at which  QW is close to symmetric, there is no splitting of the oscillations, and they correspond to doubly degenerate Landau levels. This follows from the fact that the electron density found from the oscillation period, under the assumption of twofold degeneracy of the Landau levels, coincides  with the Hall electron density within experimental accuracy. Oscillations $\delta\rho_{xx}(B)$, $\delta R_H(B)$ and $\delta C(B)$ are shown in Fig.~\ref{f5}(a). It is seen that the positions of the $\delta\rho_{xx}(B)$ and $\delta C(B)$ oscillations, as in the previously considered structures, coincide within experimental accuracy, and the amplitude of the $\delta\rho_{xx}$ oscillations is also approximately an order of magnitude greater than the amplitude of $\delta R_H(B)$.

Let us now consider the same dependences obtained for the case of a large SO splitting [Fig.~\ref{f5}(b)]. The oscillation beats clearly evident in three upper curves in Fig.~\ref{f5}(b) is manifestation of splitting of the energy spectrum of the conduction band due to the SO interaction. To separate the contributions of each of the split components, the Fourier transformation technique is used (for more details, see Ref.~\cite{Minkov22}). The results are the two lower triplets of curves in Fig.~\ref{f5}(b). As seen  the positions of the capacitance oscillations for each of the branches of the spectrum still practically coincide with the positions of the oscillations $\delta\rho_{xx}(B)$. Therewith  the phase shifts of the oscillations $\delta\rho_{xx}(B)$ and $\delta R_H(B)$ are approximately the same for both branches of the spectrum and approximately the same as for all the cases considered above.

Thus, the above results of studies of GaAs QWs  with a simple spectrum, small Zeeman and SO interactions, and HgTe QWs with a normal, inverted, and semimetal spectrum show that the positions of the oscillations $\delta\rho_{xx}(B)$ and $\delta C(B)$ in the magnetic field match. Since the oscillations of $C(B)$ are entirely determined by the oscillations of the density of states, the oscillations $\delta\rho_{xx}(B)$ are also determined by the oscillations of the density of states $\nu(B)$ in all the cases.

\section{Phases of the quantum oscillations of $\rho_{xx}(B)$ and  $R_H(B)$}
\label{sec:phase}

Let us now determine the values of the oscillation phases in $\delta\rho_{xx}(B)$ and  $\delta R_H(B)$ denoted as $\varphi_N^\rho$ and $\varphi_N^H$, respectively. As shown above for this structure (as well as for all the structures investigated in this paper) $n=\text{const}_B$.
In this case, the $\rho_{xx}$ oscillation phase is easy to determine. Indeed, the magnetic fields $B_N^\rho$ at which the minima $\rho_{xx}(B)$ are observed correspond to minima in the density of states, and they occur when the Fermi level is in the middle between the two nearest Landau levels (this does not depend on the ratio between $\hbar\omega_c$ and the broadening of the Landau levels). Hence, with taking into account the degeneracy of each Landau level in the absence of Zeeman splitting equal to  $eB/(\pi\hbar)$ one obtain that the minima of $\delta\rho_{xx}(B)$ should be observed at $B_N^\rho=2\pi\hbar n⁄(eN)$, i.e., the phase of the oscillations $\varphi_N^\rho$ must be equal to zero. The minima of $R_H(B)$ should be observed at $B_N^H=2\pi\hbar n⁄[e(N+\varphi_N^H)]$.


\begin{figure}
\includegraphics[width=0.65\linewidth,clip=true]{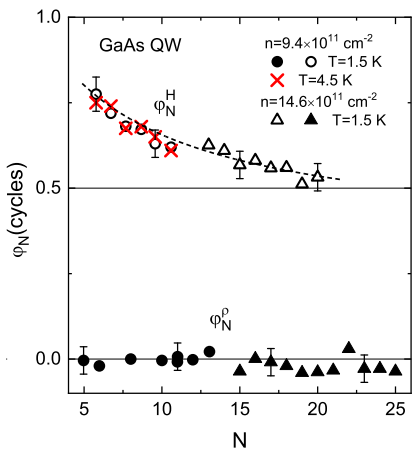}
\caption{(color online) The $N$ dependences of phases of $\rho_{xx}$ and $R_H$ oscillations, $\varphi_N^\rho$ and $\varphi_N^H$, respectively, for the structure T1520 with GaAs QW for two electron density values for $T=1.5$~K and $4.5$~K.}
\label{f6}
\end{figure}

As in the previous section, let us first consider the results for GaAs QW.
The $N$ dependences of $\varphi_N^\rho$ and $\varphi_N^H$ found in this way for two electron densities are shown in Fig.~\ref{f6}. It is seen that the oscillation phase of $\delta\rho_{xx}(B)$ is equal to zero within experimental uncertainty over the entire range of filling factors. The oscillation phase of $\delta R_H(B)$ is close to $0.5~\text{cycles}$ (or  $\pi$~radians) for large filling factors, $N>15-17$, and increases up to $0.7\pm 0.1$ for $N=5$.

Similar dependences of the oscillation phases $\varphi_N^\rho$ and $\varphi_N^H$  were obtained in GaAs QWs in Ref.~\cite{Coleridge89}. An increase in the oscillation phase $\varphi_N^H$ with decreasing $1/B_N$ was associated with  approaching  QHE regime. The role of this mechanism can be assessed experimentally. Indeed, in this case, as the temperature decreases, the deviation $\varphi_N^H$ from $0.5$ should increase and begin at larger $N$. However, in our case, as Fig.~\ref{f6} shows, such behaviour is not observed. In addition, in our range of magnetic fields, the oscillation amplitude of $\rho_{xx}$ is less than $5$~\%, so this mechanism seems unlikely.

\begin{figure}
\includegraphics[width=1.0\linewidth,clip=true]{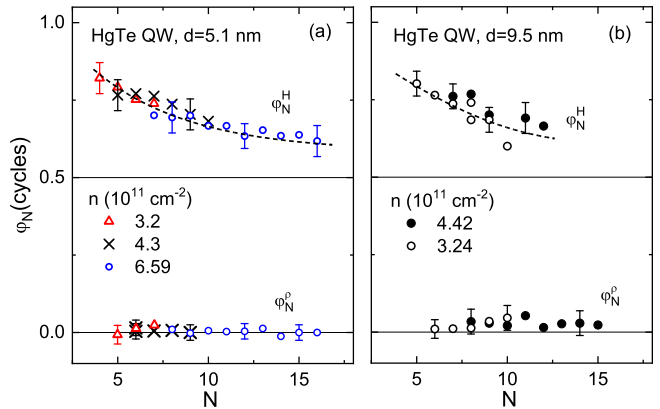}
\caption{(color online) The $N$ dependences of phases of oscillations of $\rho_{xx}$ and $R_H$ for the structures 170928 and HT71 with HgTe QWs of $5.1$~nm (a) and $9.5$~nm (b) widths for  several electron density.}
\label{f7}
\end{figure}

We turn now to results obtained for HgTe QWs.  The results of the measurements of the oscillation phases for HgTe QW with the normal spectrum are shown in Fig.~\ref{f7}(a). It is seen that all the main features are close to that for GaAs QW, namely: (i) the phase of $\rho_{xx}$ oscillations is close to zero; (ii) the $R_H$ oscillation phase somewhat decreases with increasing $N$, but remains above the theoretical value of $0.5$. Similar $N$ dependences of the oscillation phases of $\rho_{xx}$ and $R_H$ are also observed in structures with HgTe QW with an inverted spectrum, $d= 9.5$~nm [see Fig.~\ref{f7}(b)]. As for the structure 180824 with $d = 32$~nm,  it is impossible unfortunately to determine the oscillation phases   by the  method used above. This is due to the fact that only the total electron density, i.e., the sum of the electron densities in the split branches,  is constant in the magnetic field, whereas the density in each of the branches separately oscillate.

Thus, analyzing the data for GaAs and HgTe QWs in Sections~\ref{sec:RoandC} and \ref{sec:phase} we show that: (i) the oscillations of $\rho_{xx}(B)$ are entirely determined by the oscillations of the density of states; (ii) the oscillation phase of $\rho_{xx}(B)$ over the entire range of $N$ is close to zero, which corresponds to the theory, while the oscillation phase of $R_H(B)$ is close to the theoretical value of $0.5$~cycles only for large $N$ values and increases as they decrease; (iii) Such the increase is not associated with approaching the QHE regime; (iv) Physical  reason for the growth of oscillation phase of  $\delta R_H(B)$  observed in many papers \cite{Coleridge89,Simon94,Mani09,Qian17} remains unclear.

\section{Amplitudes of the quantum oscillations of $\rho_{xx}(B)$ and  $R_H(B)$}
\label{sec:ampl}

As discussed in Section~\ref{sec:intr}, the theory predicts a simple ratio of the oscillation amplitudes $\delta\rho_{xx}$ and $\delta R_H$, Eq.~(\ref{eq100}).  Let us compare the experimental values of
$$R=-2(\mu B)^2A_H/A_\rho$$
with the theoretical values for the structures considered above.

As in Sections~ \ref{sec:RoandC} and \ref{sec:phase}, we first consider the results of measurements for  GaAs QW with a simple spectrum, in which the Zeeman and spin-orbit splitting is negligibly small in the actual range of magnetic fields,  and electron densities.

\begin{figure}
\includegraphics[width=1.0\linewidth,clip=true]{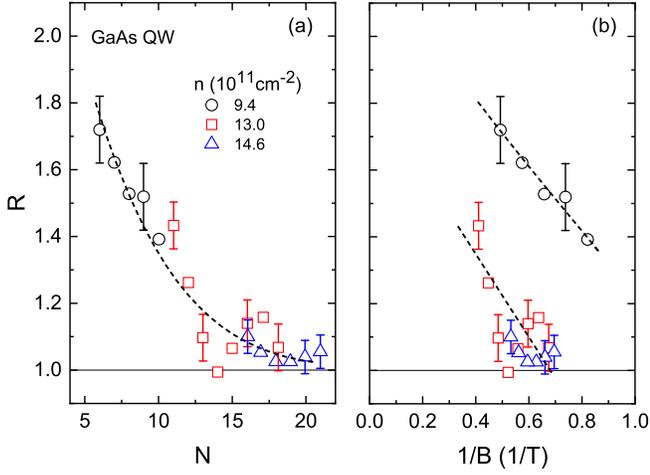}
\caption{(color online) The  $R$ value plotted as function of the  filling factor (a) and reciprocal magnetic field (b) for the structure T1520 with GaAs QW for three electron density values, $T=1.4$~K. }
\label{f8}
\end{figure}

The $N$ dependences of $R$   plotted in Fig.~\ref{f8} for three electron densities show that the $R$ value is close to unity for filling factors $N>13-14$ in accordance with Eq.~(\ref{eq100}). As $N$ decreases, the ratio slightly increases. Note that ratio deviates from theory at about the same $N$ as the  dependence $\varphi_N^H(N)$ deviates from $0.5$ (see Fig.~\ref{f6}). This result also agrees with the results from Ref.~\cite{Coleridge89}. Note,  the $R$ vs $N$ dependences obtained for  different electron densities lie on the same curve, while the $R$ vs $1/B$ dependences do not.

\begin{figure}
\includegraphics[width=0.9\linewidth,clip=true]{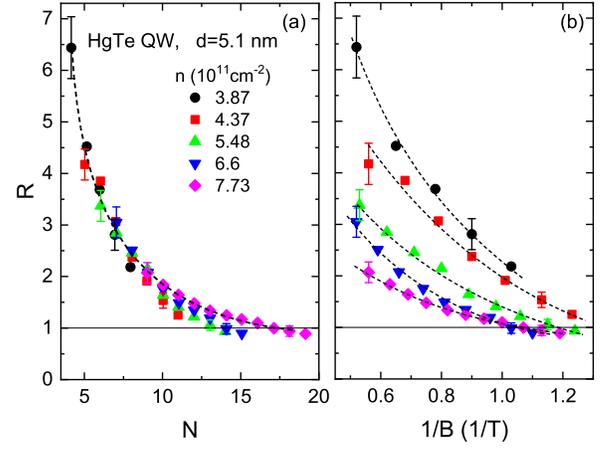}
\caption{(color online) The $R$ values plotted against the filling factor  (a) and reciprocal magnetic field (b) for  the structure 170938 with HgTe QW of  $5.1$~nm width with normal spectrum for different electron densities.  $T=3.7$~K.}
\label{f9}
\end{figure}

Let us now consider the $R$ versus $N$ dependence  for HgTe QW. First, as in the previous section, we consider the results for the structures 170928  with a normal spectrum, $d=5.1$~nm (Fig.~\ref{f9}). It is seen that they are qualitatively similar to those for GaAs QW (Fig.~\ref{f8}); the $R$ vs $N$ dependences obtained for different $n$ lie on the common curve, while the $R$ vs $1/B$ dependences differ significantly.  Along with this there is a quantitative difference with GaAs QW. As seen from Fig.~\ref{f9}(b) the $R$ value being close to unity at $N>11-12$ increases very strongly with decreasing $N$ and  becomes three times larger than that in GaAs QW at $N=6$ [see Fig.~\ref{f8}(b)].

\begin{figure}
\includegraphics[width=1.0\linewidth,clip=true]{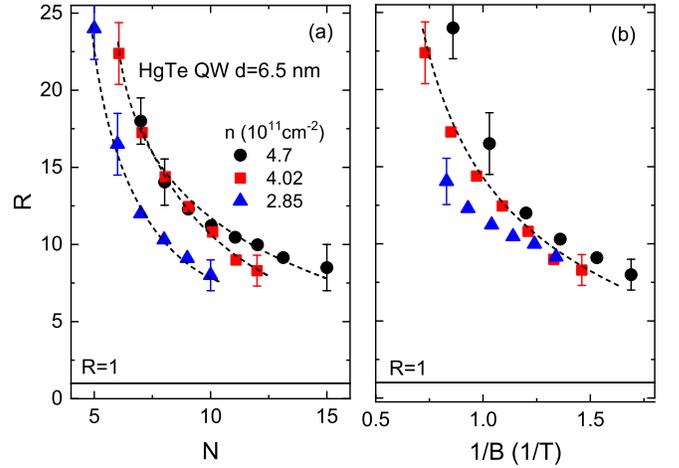}
\caption{(color online) The $R$ values plotted against the filling factor (a) and reciprocal magnetic field (b) for  the structure H1023 with HgTe QW with inverted spectrum, $d=6.5$~nm, for different electron densities.  $T=3.7$~K.  }
\label{f10}
\end{figure}

We now inspect the results for the structure H1023 with an inverted spectrum with $d=6.5$~nm that is a little more than $d_c=6.3$~nm\footnote{The inverted character of the spectrum in this structure is justified in our article \cite{Minkov16}, in which the intersection of Landau levels belonging to the conduction and valence bands, specific for quantum wells with an inverted spectrum, was observed.}.  Fig.~\ref{f10} shows that the $R$~vs~$N$ and $R$~vs~$1/B$ dependences  differ significantly from these observed in the structure with a normal spectrum shown in Fig.~\ref{f9}. The $R$ value  reaches $24$  at $N=6$. With increasing $N$, the ratio decreases to a value of $6-8$, which is much larger than the theoretical value equal to unity.
The $N$ dependence as well as the $1/B$ dependence of the ratio found for different electron densities do not fall on the one curve.

\begin{figure}
\includegraphics[width=1.0\linewidth,clip=true]{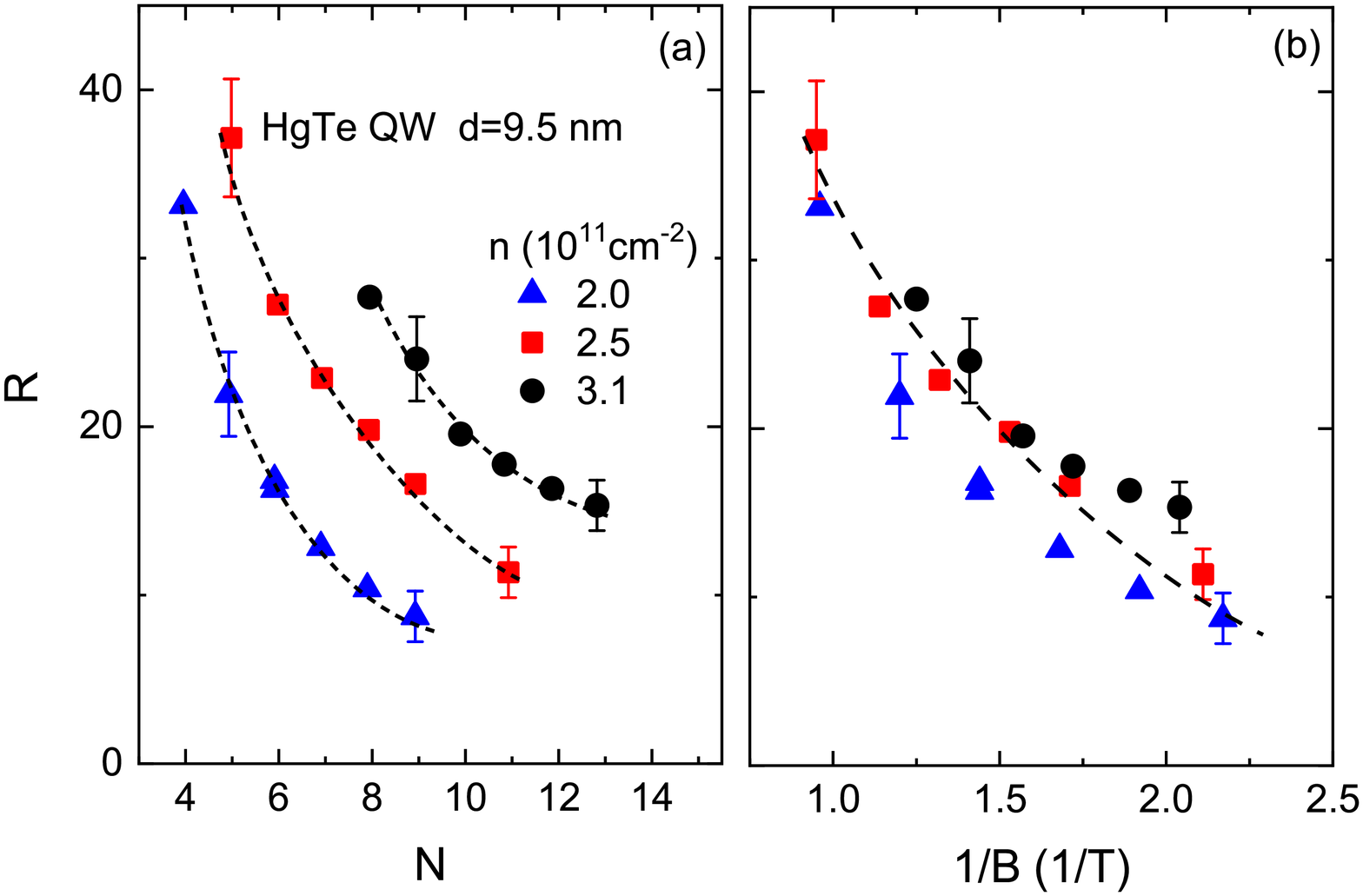}
\caption{(color online) The $R$ values plotted against the filling factor (a) and reciprocal magnetic field (b) for the structure HT71 with  HgTe QW with inverted spectrum, $d=9.5$~nm, for different electron densities.  $T=3.7$~K.  }
\label{f11}
\end{figure}

Finally, let us consider the results for a structure HT71 with $d=9.5$~nm which is significantly larger than $d_c$. Fig.~\ref{f11} shows that in this case the ratio is even larger than that in the structure H1023 with $d = 6.5$~nm for all available filling factors. In contrast to the structure 170928 with a normal spectrum, in this case, the ratio plotted as function of $1/B$ for different electron densities falls  on one curve, whereas the $R$ values plotted as function of $N$ does not.

Thus, in contrast to the phases of the $\rho_{xx}$ and $R_H$ oscillations, whose dependences on the filling factor are practically the same in GaAs and HgTe quantum wells both with normal and inverted spectra, the dependences of the ratio of  amplitude of the oscillation of $\rho_{xx}(B)$ and $R_H(B)$ on the filling factor and on reciprocal magnetic field differ greatly. In  GaAs QWs, the  ratio for large $N$ is close to unity, as predicted by the theory and increases somewhat with decreasing $N$.  In HgTe  QWs, the  ratio increases strongly with decreasing $N$ and differently in structures with normal and inverted spectra. In the narrowest QW the $R$ versus $N$ dependences  lie on the same curve for the  different electron density, while in the widest QW the dependences the $R$ versus $1/B$ lie on the same curve.

\section{Discussion}
\label{sec:disc}

Let us discussed the differences between GaAs and HgTe QWs, which could  be responsible for the different behaviour of the effects considered in the previous Sections.

The obvious difference is the nonparabolicity of the energy spectrum. In GaAs QWs, the nonparabolicity is very low in the actual range of electron densities, while in HgTe QWs it is much stronger. But for large $N$, the position of the Landau levels in the magnetic field at a fixed electron density or the Fermi energy does not depend on the nonparabolicity  of the energy spectrum.

The other difference is the strength of SO interaction of the Rashba type. In GaAs QWs, the SO interaction  is weak. It is much stronger in HgTe QWs. When the SO interaction is less than the broadening of the Landau levels and $\hbar\omega_c$ within actual magnetic field range,  the positions of the maxima of the density of states  is the same as in the absence of SO splitting. But in this case, the SO interaction can change the scattering due to the SO interaction with the potential of the scatterers.
We are not aware of any theoretical papers in which the effect of this interaction on the amplitudes of  $\delta \rho_{xx}$ and $\delta R_H$ oscillations was considered.

Luckily, the role of this interaction can be tested experimentally by comparing the amplitudes of the $\delta \rho_{xx}$ and $\delta R_H$ oscillations in structures with a close energy spectrum, but with different Rashba parameters. In our case, these are GaAs and In$_x$Ga$_{1-x}$As QWs. In the In$_x$Ga$_{1-x}$As QWs, the Rashba parameter is much larger than that in GaAs QWs, whereas  the other parameters of the spectrum are close for both types of the QWs if the indium concentration $x$ is relatively small. So, for $x=0.2$ they are only $(10-15)$\% smaller than that in GaAs QWs (see Table~\ref{tab2}).
\begin{table*}
\caption{The parameters of  structures T1520 and Z76}
\label{tab2}
\begin{ruledtabular}
\begin{tabular}{cccccccc}
$V_g(V)$ & $n$(cm$^{-2})$ &  $\mu$ (m$^2$/Vs) & $m/m_0$ & $\Delta_L$ (meV)& $\tau/\tau_q$\\
\colrule
structure T1520 &  &  & &  & \\
\colrule
 -2.5 & $9.40\times 10^{11}$ & $0.89$   &  & $3.20$ &   $3.54$ \\
  -1.0 & $1.30\times 10^{12}$ & $1.50$   &$0.067$  &  $2.76$ &  $5.06$  \\
 -0.25 & $1.46\times 10^{12}$ & $1.83$   &  &  $2.56$ &  $5.6$  \\
 \colrule
structure Z76 &  &  & &  & \\
\colrule
 -0.70 & $(5.16\pm 0.02)\times 10^{11}$ & $2.45\pm 0.02$   &  & $1.90\pm 0.05$ &   $5.42$ \\
  -0.45 & $5.48\times 10^{11}$ & $2.56$   &$0.06\pm 0.005$  &  $1.64$ &  $4.96$  \\
 -0.20 & $5.90\times 10^{11}$ & $2.78$   &  &  $1.25$ &  $4.15$  \\
\end{tabular}
\end{ruledtabular}
\footnotetext[1]{For $V_g=0$~V}
\end{table*}

\begin{figure}
\includegraphics[width=0.7\linewidth,clip=true]{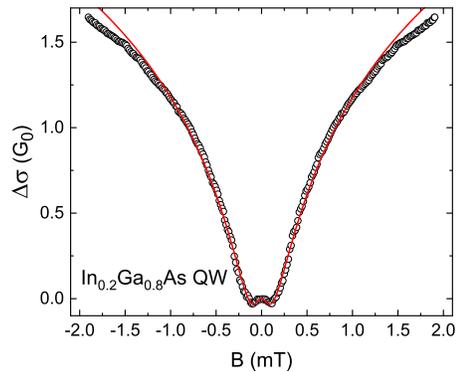}
\caption{(color online) The antilocalization magnetoconductivity in the structure Z76 with the In$_{0.2}$Ga$_{0.8}$As QW for $n=6.5\times 10^{11}$~cm$^{-2}$, $\mu=2.58$~m$^2$/Vs.  The symbols are the data, the curve is the result of the best fit  with the phase and spin relaxation times equal to $1.04\times 10^{-10}$~s and  $3.6\times 10^{-11}$~s, respectively (for more details, see Ref.~\cite{Min05}). $T=0.47$~K.}
\label{f13}
\end{figure}

Although the SO interaction in In$_{0.2}$Ga$_{0.8}$As QWs is stronger than that un GaAs QWs, it is relatively weak  to manifest itself in the SdH oscillations and in their Fourier spectrum. In this case the presence and the  value of the SO splitting  can be estimated from the magnetic-field behaviour of the weak localization correction to the conductivity. As known the SO interaction  leads to antilocalization behaviour of the magnetoresistance in low magnetic fields. The dependence $\Delta \sigma(B)= \rho_{xx}^{-1}(B)-\rho_{xx}^{-1}(0)$ for the heterostructure Z76 with the  In$_{0.2}$Ga$_{0.8}$As QW shown in Fig.~\ref{f13} demonstrates clearly the antilocalization behaviour in low fields. The processing of these results similar to that used in \cite{Min05} gives the value of the SO splitting of about ($0.10-0.15$)~meV.

As for the structure T1520, only weak localization behaviour has been observed. This means that the value of the SO splitting in the GaAs QW is much smaller than that in In$_{0.2}$Ga$_{0.8}$As QW.

Thus the only factor that differs significantly in these structures is the strength of the SO  interaction.

\begin{figure}
\includegraphics[width=0.9\linewidth,clip=true]{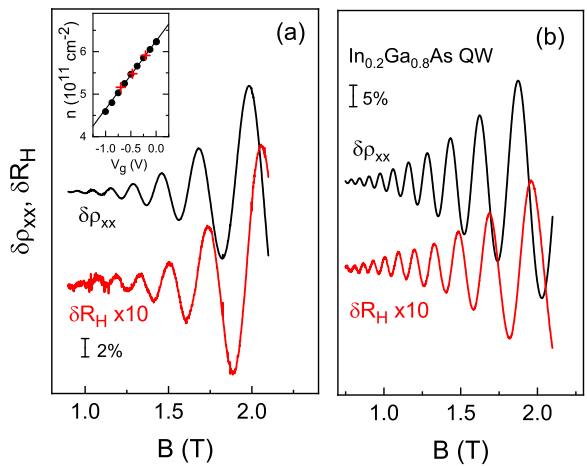}
\caption{(color online) Oscillating parts $\delta\rho_{xx}(B)$ and $\delta R_H(B)$ in the structure Z76 with In$_{0.2}$Ga$_{0.8}$As QW for two electron densities $n=5.16\times 10^{11}$~cm$^{-2}$ (a) and $5.91\times 10^{11}$~cm$^{-2}$ (b). The inset in panel (a) shows the gate voltage dependence of electron density found from the Hall effect (circles) and from SdH oscillations (crosses). The line in the inset is the dependence $n=(6.25+1.6\,V_g)\times 10^{11}$~cm$^{-2}$. $T= 1.7$~K.
}
\label{f14}
\end{figure}

Let us consider the results of measurement of SdH oscillations obtained for the structure Z76 with In$_{0.2}$Ga$_{0.8}$As QW. The gate voltage dependences of the electron density, plotted in the inset in Fig.~\ref{f14} shows that, just as in the GaAs structure, the electron density  found from the Hall effect and the SdH oscillations coincide and depend linearly on $V_g$.

\begin{figure}
\includegraphics[width=1.0\linewidth,clip=true]{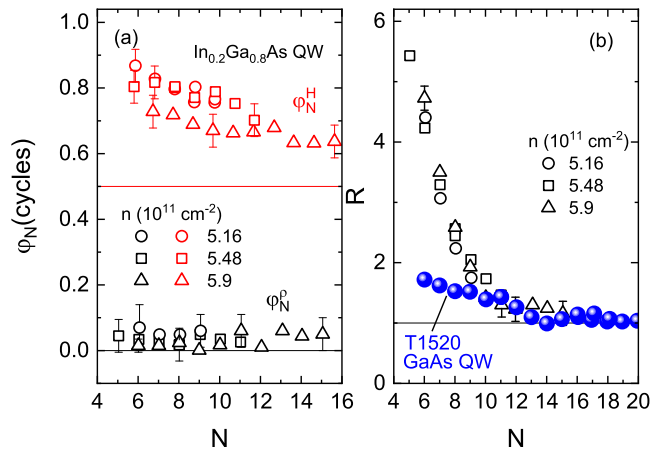}
\caption{(color online) (a) Dependences of phases of oscillations of $\rho_{xx}(B)$ and $R_H(B)$ on $N$.  (b) The $R$ values plotted against the filling factor.  The structure Z76 with In$_{0.2}$Ga$_{0.8}$As QW, $T= 1.7$~K. The balls in the panel (b) are the data for GaAs QW.
}
\label{f15}
\end{figure}

The $N$ dependences of the oscillation phases of $\delta\rho_{xx}$ and $\delta R_H$  for the structure Z76  shown in Fig.~\ref{f15}(a) practically coincide with those for the structure T1520 (Fig.~\ref{f6}). As for the $N$ dependences of the  ratio of the oscillation amplitudes, they are  qualitatively similar, but as Fig.~\ref{f15}(b) shows $R(N)$  for the structure Z76 exhibits much stronger increase of $R$ with decreasing $N$. A very similar behaviour of $R(N)$ is observed for the structure 170938 with HgTe QW of  $5.1$~nm width  [see Fig.~\ref{f9}].

All the above results make us think that the   difference in the behaviour of $R(N)$ is  a consequence of a strong difference of strength of SO interaction. As discussed above, the amplitudes of the $\rho_{xx}$ and $R_H$ oscillations are determined by the oscillations of the density of states, and as long as the spin-orbit splitting remains less than the broadening of the Landau levels, the SO interaction should not change the ratio of the amplitudes of the $\rho_{xx}$ and $R_H$ oscillations.

One can assume that the  SO interaction with a scatterer potential changes the scattering in such a way that leads to the observed behaviour of $R(N)$.

As far as we understand it is the SO interaction  with the potential of scatterers that is the cause of the spin Hall effect.

We are aware of only one paper \cite{Dyakonov08} which provides a phenomenological description of the spin Hall effect in a magnetic field.
This consideration was made for the three-dimensional (3D) case. Nevertheless, $\rho_{xx}$ and $\rho_{xy}$ estimates made within this approach for the 2D case  gives the $N$ dependence of ratio $R$ qualitatively similar to the experiment; its value  increases with decreasing $N$. For a reliable conclusion that all the above results on difference in behavior  $R(N)$ and $R(1/B)$ with the increasing HgTe-QW width requires more detailed theoretical calculations taking into account, among other things, the features of the spectrum of HgTe QW.

\section{Conclusion}
\label{sec:concl}
In this paper, systematic experimental studies of the oscillations of  $\rho_{xx}$, $R_H$ and capacitance both for GaAs, and In$_x$Ga$_{1-x}$As structures with a simple spectrum of the conduction band and HgTe structures with a nonparabolic complicated spectrum and strong spin-orbit interaction were carried out. The HgTe structures with normal and inverted spectra, $d<6.3$~nm and $d>6.3$~nm, respectively, were studied.

It has been found that the positions of the $\rho_{xx}$ and $C$ oscillations in a magnetic field  coincide for all the structures under study. This indicates that the oscillations of $\rho_{xx}$ is determined only by the oscillations of the density of states in the magnetic field.

For all the structures, the phase difference in the $\rho_{xx}$ and $R_H$ oscillations  agrees with the theoretical value $\pi$ only for large filling factors ($N>15$) and increases at smaller $N$. Such behavior is consistent with the results of Ref.~\cite{Coleridge89}, in which it is assumed that the phase difference  at $N<15$ is due to approaching the QHE regime. However, our measurements performed for different temperatures shows that this is not the case.  So, the physical reason for the difference between experiment and theory remains unclear.

The most surprising result relates to the ratio of the amplitudes of the $\rho_{xx}$ and $R_H$ oscillations $A_\rho/A_H$. In the structure with GaAs QW with a simple spectrum and a negligible Zeeman and spin-orbit splitting, the ratio of the amplitudes $A_\rho/A_H$ is close to the theoretical $-2\mu^2 B^2$ (but also at $N>15$). In contrast, in structures with HgTe QW, this ratio differs from theoretical predictions and differently in structures with normal and inverted spectrum. In structures with a normal spectrum, it tends to the theoretical value with the growth of $N$. In structures with an inverted spectrum, it remains different from the theoretical one for all available $N$ and does not tend to the theoretical one with increasing $N$.

It is assumed that such a difference in the ratio of amplitudes $A_\rho/A_H$ relates not to the features of the energy spectrum of HgTe QW, but to the features of electron scattering due to SO interaction with the potential of the scatterers.

To elucidate the role of this mechanism, all the same measurements were carried out for the structure with In$_{0.2}$Ga$_{0.8}$As, which spectrum  is very close to that of  GaAs QW, but differs only by a significantly stronger spin-orbit interaction. This is evidenced by measurements of weak antilocalization magnetoresistivity. Its analysis allows us to determine the value of spin-orbit splitting as $(0.1-0.15)$~meV, which is significantly less than the broadening of the Landau levels $\Delta_L\simeq 1.9$~meV. The positions of the $\rho_{xx}$ and $C$ oscillations, the phase difference of $\rho_{xx}$ and $R_H$ and its dependence on the filling factor in this structure are close to those observed in GaAs QW. But the ratio of the oscillation amplitudes $A_\rho/A_H$  differs strongly. It and its $N$ dependence  become close to that observed in HgTe QW with normal spectrum.

Thus, we believe that the change in scattering due to the SO interaction with the potential of the scatterers is the reason for the difference in the ratio of the amplitudes $A_\rho/A_H$  in the HgTe quantum well. Such a change in scattering due to SO interaction with the potential of the scatterers is one of the reasons for the spin-Hall effect. There are many papers on the spin-Hall effect in the absence of a magnetic field \cite{Sinova15}.
We know only one paper \cite{Dyakonov08} in which the phenomenological model of the spin-Hall effect in a magnetic field is considered for 3D systems. Estimates for 2D systems performed along this line of attack give qualitative agreement with experimental results. For a reliable conclusion that all of the above leads to a difference in the behaviour of $R(N)$ and $R(1/B)$
with an increase in the width of HgTe QW, more detailed theoretical calculations are required, taking into account, among other things, the features of the QW spectrum.

If the peculiarity in the behaviour of the ratio of the amplitudes of the $\rho_{xx}$ and $R_H$ oscillations in the magnetic field is indeed related to the spin-orbit interaction with the potential of the scatterers, then this would give a purely electrical method for measuring the spin-Hall effect.

\begin{acknowledgments}
We are grateful to I. Gornyi for useful discussions. The research was supported by the Ministry of Science and Higher Education of the Russian Federation under projects Nos. 075-15-2020-797 (13.1902.21.0024) and FEUZ-2023-0017.
\end{acknowledgments}


\begin{thebibliography}{29}%
\makeatletter
\providecommand \@ifxundefined [1]{%
 \@ifx{#1\undefined}
}%
\providecommand \@ifnum [1]{%
 \ifnum #1\expandafter \@firstoftwo
 \else \expandafter \@secondoftwo
 \fi
}%
\providecommand \@ifx [1]{%
 \ifx #1\expandafter \@firstoftwo
 \else \expandafter \@secondoftwo
 \fi
}%
\providecommand \natexlab [1]{#1}%
\providecommand \enquote  [1]{``#1''}%
\providecommand \bibnamefont  [1]{#1}%
\providecommand \bibfnamefont [1]{#1}%
\providecommand \citenamefont [1]{#1}%
\providecommand \href@noop [0]{\@secondoftwo}%
\providecommand \href [0]{\begingroup \@sanitize@url \@href}%
\providecommand \@href[1]{\@@startlink{#1}\@@href}%
\providecommand \@@href[1]{\endgroup#1\@@endlink}%
\providecommand \@sanitize@url [0]{\catcode `\\12\catcode `\$12\catcode
  `\&12\catcode `\#12\catcode `\^12\catcode `\_12\catcode `\%12\relax}%
\providecommand \@@startlink[1]{}%
\providecommand \@@endlink[0]{}%
\providecommand \url  [0]{\begingroup\@sanitize@url \@url }%
\providecommand \@url [1]{\endgroup\@href {#1}{\urlprefix }}%
\providecommand \urlprefix  [0]{URL }%
\providecommand \Eprint [0]{\href }%
\providecommand \doibase [0]{https://doi.org/}%
\providecommand \selectlanguage [0]{\@gobble}%
\providecommand \bibinfo  [0]{\@secondoftwo}%
\providecommand \bibfield  [0]{\@secondoftwo}%
\providecommand \translation [1]{[#1]}%
\providecommand \BibitemOpen [0]{}%
\providecommand \bibitemStop [0]{}%
\providecommand \bibitemNoStop [0]{.\EOS\space}%
\providecommand \EOS [0]{\spacefactor3000\relax}%
\providecommand \BibitemShut  [1]{\csname bibitem#1\endcsname}%
\let\auto@bib@innerbib\@empty
\bibitem [{\citenamefont {Lifshits}\ and\ \citenamefont
  {Kosevich}(1955)}]{LifKos55}%
  \BibitemOpen
  \bibfield  {author} {\bibinfo {author} {\bibfnamefont {I.~M.}\ \bibnamefont
  {Lifshits}}\ and\ \bibinfo {author} {\bibfnamefont {A.~M.}\ \bibnamefont
  {Kosevich}},\ }\bibfield  {title} {\bibinfo {title} {Theory of magnetic
  susceptibility in metals at low temperatures},\ }\href@noop {} {\bibfield
  {journal} {\bibinfo  {journal} {Zh. Eksp. Teor. Fiz.}\ }\textbf {\bibinfo
  {volume} {29}},\ \bibinfo {pages} {730} (\bibinfo {year} {1955})},\
  \translation{Sov. Phys. JETP \textbf{2}, 636 (1956)}\BibitemShut {NoStop}%
\bibitem [{\citenamefont {Ando}(1974)}]{Ando74}%
  \BibitemOpen
  \bibfield  {author} {\bibinfo {author} {\bibfnamefont {T.}~\bibnamefont
  {Ando}},\ }\bibfield  {title} {\bibinfo {title} {Theory of quantum transport
  in a two-dimensional electron system under magnetic fields. iv. oscillatory
  conductivity},\ }\href@noop {} {\bibfield  {journal} {\bibinfo  {journal} {J
  . Phys. Soc. Japan}\ }\textbf {\bibinfo {volume} {37}},\ \bibinfo {pages}
  {1233} (\bibinfo {year} {1974})}\BibitemShut {NoStop}%
\bibitem [{\citenamefont {Ando}\ \emph {et~al.}(1975)\citenamefont {Ando},
  \citenamefont {Matsumoto},\ and\ \citenamefont {Uemura}}]{Ando75}%
  \BibitemOpen
  \bibfield  {author} {\bibinfo {author} {\bibfnamefont {T.}~\bibnamefont
  {Ando}}, \bibinfo {author} {\bibfnamefont {T.}~\bibnamefont {Matsumoto}},\
  and\ \bibinfo {author} {\bibfnamefont {Y.}~\bibnamefont {Uemura}},\
  }\href@noop {} {\bibfield  {journal} {\bibinfo  {journal} {J . Phys. Soc.
  Japan}\ }\textbf {\bibinfo {volume} {39}},\ \bibinfo {pages} {279} (\bibinfo
  {year} {1975})}\BibitemShut {NoStop}%
\bibitem [{\citenamefont {A.~Isihara}\ and\ \citenamefont
  {Smr\v{e}cka}(1986)}]{Isihara86}%
  \BibitemOpen
  \bibfield  {author} {\bibinfo {author} {\bibfnamefont {A.}~\bibnamefont
  {A.~Isihara}}\ and\ \bibinfo {author} {\bibfnamefont {L.}~\bibnamefont
  {Smr\v{e}cka}},\ }\href@noop {} {\bibfield  {journal} {\bibinfo  {journal}
  {J. Phys. C: Solid State Phys}\ }\textbf {\bibinfo {volume} {19}},\ \bibinfo
  {pages} {6777} (\bibinfo {year} {1986})}\BibitemShut {NoStop}%
\bibitem [{\citenamefont {Novokshonov}(2013)}]{Novokshonov13}%
  \BibitemOpen
  \bibfield  {author} {\bibinfo {author} {\bibfnamefont {S.}~\bibnamefont
  {Novokshonov}},\ }\bibfield  {title} {\bibinfo {title} {Magneto-intersubband
  oscillations of the kinetic coefficients of two-dimensional system with a
  spin-orbit interaction},\ }\href {https://doi.org/10.1063/1.4803177}
  {\bibfield  {journal} {\bibinfo  {journal} {Low Temperature Physics}\
  }\textbf {\bibinfo {volume} {39}} (\bibinfo {year} {2013})}\BibitemShut
  {NoStop}%
\bibitem [{\citenamefont {Dmitriev}\ \emph {et~al.}(2012)\citenamefont
  {Dmitriev}, \citenamefont {Mirlin}, \citenamefont {Polyakov},\ and\
  \citenamefont {Zudov}}]{Dmitriev12}%
  \BibitemOpen
  \bibfield  {author} {\bibinfo {author} {\bibfnamefont {I.~A.}\ \bibnamefont
  {Dmitriev}}, \bibinfo {author} {\bibfnamefont {A.~D.}\ \bibnamefont
  {Mirlin}}, \bibinfo {author} {\bibfnamefont {D.~G.}\ \bibnamefont
  {Polyakov}},\ and\ \bibinfo {author} {\bibfnamefont {M.~A.}\ \bibnamefont
  {Zudov}},\ }\bibfield  {title} {\bibinfo {title} {Nonequilibrium phenomena in
  high landau levels},\ }\href {https://doi.org/10.1103/RevModPhys.84.1709}
  {\bibfield  {journal} {\bibinfo  {journal} {Rev. Mod. Phys.}\ }\textbf
  {\bibinfo {volume} {84}},\ \bibinfo {pages} {1709} (\bibinfo {year}
  {2012})}\BibitemShut {NoStop}%
\bibitem [{Note1()}]{Note1}%
  \BibitemOpen
  \bibinfo {note} {As seen from Eq.~(\ref {eq80}), the magnetic field of the
  $N$th $\rho _{xx}$ minimum ($B_N$) corresponds to the minimum of the density
  of states at the Fermi level, when the integer number of Landau levels are
  occupied, i.e., the Fermi level is located between the two Landau levels.
  Note, the values of $B_N$ are different for the two regimes: $E_F=\protect
  \text {const}_B$ and $n=\protect \text {const}_B$. In the first case, taking
  into account the fact that the energy of the Landau levels is $E_N=\hbar
  \omega _c(N+1/2+\phi )$, where $\phi $ is the Berry phase, the fields of
  $\rho _{xx}$ minima are determined by the condition $\varepsilon _F=E_N+\hbar
  \omega _c/2= e\hbar B_N/m\times (N+1/2+1/2+\phi )$. Thus we get
  $B_N=\varepsilon _F/[e\hbar /m (N+1+\phi )]$, i.e., an extrapolation of the
  $N$ versus $1/B_N$ experimental plot allows us to obtain the Barry phase. In
  the second case, $n=\protect \text {const}_B$, the condition for filling an
  integer number of Landau levels is simpler: the number of states at twofold
  degenerate Landau levels in the field $B$ is $eB/(\pi \hbar )$, so $B_N=\pi
  \hbar n/(eN)$, This shows that at $n=\protect \text {const}_B$ $B_N$ does not
  include the Berry phase \cite {Kuntsevich18}.}\BibitemShut {Stop}%
\bibitem [{\citenamefont {Coleridge}\ \emph {et~al.}(1989)\citenamefont
  {Coleridge}, \citenamefont {Stoner},\ and\ \citenamefont
  {Fletcher}}]{Coleridge89}%
  \BibitemOpen
  \bibfield  {author} {\bibinfo {author} {\bibfnamefont {P.~T.}\ \bibnamefont
  {Coleridge}}, \bibinfo {author} {\bibfnamefont {R.}~\bibnamefont {Stoner}},\
  and\ \bibinfo {author} {\bibfnamefont {R.}~\bibnamefont {Fletcher}},\
  }\bibfield  {title} {\bibinfo {title} {Low-field transport coefficients in
  gaas/${\mathrm{ga}}_{1\mathrm{\ensuremath{-}}\mathrm{x}}$${\mathrm{al}}_{\mathrm{x}}$as
  heterostructures},\ }\href {https://doi.org/10.1103/PhysRevB.39.1120}
  {\bibfield  {journal} {\bibinfo  {journal} {Phys. Rev. B}\ }\textbf {\bibinfo
  {volume} {39}},\ \bibinfo {pages} {1120} (\bibinfo {year}
  {1989})}\BibitemShut {NoStop}%
\bibitem [{\citenamefont {Mikhailov}\ \emph {et~al.}(2006)\citenamefont
  {Mikhailov}, \citenamefont {Smirnov}, \citenamefont {Dvoretsky},
  \citenamefont {Sidorov}, \citenamefont {Shvets}, \citenamefont {Spesivtsev},\
  and\ \citenamefont {Rykhlitski}}]{Mikhailov06}%
  \BibitemOpen
  \bibfield  {author} {\bibinfo {author} {\bibfnamefont {N.~N.}\ \bibnamefont
  {Mikhailov}}, \bibinfo {author} {\bibfnamefont {R.~N.}\ \bibnamefont
  {Smirnov}}, \bibinfo {author} {\bibfnamefont {S.~A.}\ \bibnamefont
  {Dvoretsky}}, \bibinfo {author} {\bibfnamefont {Y.~G.}\ \bibnamefont
  {Sidorov}}, \bibinfo {author} {\bibfnamefont {V.~A.}\ \bibnamefont {Shvets}},
  \bibinfo {author} {\bibfnamefont {E.~V.}\ \bibnamefont {Spesivtsev}},\ and\
  \bibinfo {author} {\bibfnamefont {S.~V.}\ \bibnamefont {Rykhlitski}},\
  }\bibfield  {title} {\bibinfo {title} {Growth of hgcdte nanostructures by
  molecular beam epitaxy with ellipsometric control},\ }\href
  {https://doi.org/10.1504/IJNT.2006.008725} {\bibfield  {journal} {\bibinfo
  {journal} {Int. J. Nanotechnology}\ }\textbf {\bibinfo {volume} {3}},\
  \bibinfo {pages} {120 } (\bibinfo {year} {2006})}\BibitemShut {NoStop}%
\bibitem [{Note2()}]{Note2}%
  \BibitemOpen
  \bibinfo {note} {In the equations given in Section~\ref {sec:intr}, only the
  short-range potential scattering is taken into account, so there are no
  monotonous parts of the dependences $\rho _{xx}(B)$, $R_H(B)$. In real
  structures, there are various mechanisms that can lead to their appearance,
  for instance, the scattering by potential fluctuations, the electron-electron
  interaction, etc.}\BibitemShut {Stop}%
\bibitem [{Note3()}]{Note3}%
  \BibitemOpen
  \bibinfo {note} {As shown in numerous papers (see, e.g., \cite
  {Gerchikov90,Zhang01,Novik05,Bernevig06,ZholudevPhD,Ren2016} and references
  therein) different types of spectrum are realized depending on the QW width.
  When QW is narrow, $d < d_c$, the ordering of energy subbands of spatial
  quantization is analogous to that in conventional semiconductors; the highest
  valence subband at $k = 0$ is formed from the heavy hole $\Gamma _8$ states,
  while the lowest conduction subband is formed both from the $\Gamma _6$
  states and light hole $\Gamma _8$ states. For a thicker HgTe layer, $d >
  d_c$, the quantum well is in the inverted regime; the lowest conduction
  subband is formed from the heavy hole $\Gamma _8$ states \cite {Dyak82e},
  whereas the subband formed from the $\Gamma _6$ states and light hole $\Gamma
  _8$ states sinks into the valence band. In the inverted regime, the spectrum
  becomes semimetallic at $d\gtrsim 15$~nm due to overlapping of the valence
  and conduction bands.}\BibitemShut {Stop}%
\bibitem [{\citenamefont {Minkov}\ \emph {et~al.}(2020)\citenamefont {Minkov},
  \citenamefont {Rut}, \citenamefont {Sherstobitov}, \citenamefont {Dvoretski},
  \citenamefont {Mikhailov}, \citenamefont {Solov'ev}, \citenamefont {Chernov},
  \citenamefont {Ivanov},\ and\ \citenamefont {Germanenko}}]{Minkov20-2}%
  \BibitemOpen
  \bibfield  {author} {\bibinfo {author} {\bibfnamefont {G.~M.}\ \bibnamefont
  {Minkov}}, \bibinfo {author} {\bibfnamefont {O.~E.}\ \bibnamefont {Rut}},
  \bibinfo {author} {\bibfnamefont {A.~A.}\ \bibnamefont {Sherstobitov}},
  \bibinfo {author} {\bibfnamefont {S.~A.}\ \bibnamefont {Dvoretski}}, \bibinfo
  {author} {\bibfnamefont {N.~N.}\ \bibnamefont {Mikhailov}}, \bibinfo {author}
  {\bibfnamefont {V.~A.}\ \bibnamefont {Solov'ev}}, \bibinfo {author}
  {\bibfnamefont {M.~Y.}\ \bibnamefont {Chernov}}, \bibinfo {author}
  {\bibfnamefont {S.~V.}\ \bibnamefont {Ivanov}},\ and\ \bibinfo {author}
  {\bibfnamefont {A.~V.}\ \bibnamefont {Germanenko}},\ }\bibfield  {title}
  {\bibinfo {title} {Magneto-intersubband oscillations in two-dimensional
  systems with an energy spectrum split due to spin-orbit interaction},\ }\href
  {https://doi.org/10.1103/PhysRevB.101.245303} {\bibfield  {journal} {\bibinfo
   {journal} {Phys. Rev. B}\ }\textbf {\bibinfo {volume} {101}},\ \bibinfo
  {pages} {245303} (\bibinfo {year} {2020})}\BibitemShut {NoStop}%
\bibitem [{\citenamefont {Minkov}\ \emph {et~al.}(2022)\citenamefont {Minkov},
  \citenamefont {Aleshkin}, \citenamefont {Rut}, \citenamefont {Sherstobitov},
  \citenamefont {Dvoretski}, \citenamefont {Mikhailov},\ and\ \citenamefont
  {Germanenko}}]{Minkov22}%
  \BibitemOpen
  \bibfield  {author} {\bibinfo {author} {\bibfnamefont {G.~M.}\ \bibnamefont
  {Minkov}}, \bibinfo {author} {\bibfnamefont {V.~Y.}\ \bibnamefont
  {Aleshkin}}, \bibinfo {author} {\bibfnamefont {O.~E.}\ \bibnamefont {Rut}},
  \bibinfo {author} {\bibfnamefont {A.~A.}\ \bibnamefont {Sherstobitov}},
  \bibinfo {author} {\bibfnamefont {S.~A.}\ \bibnamefont {Dvoretski}}, \bibinfo
  {author} {\bibfnamefont {N.~N.}\ \bibnamefont {Mikhailov}},\ and\ \bibinfo
  {author} {\bibfnamefont {A.~V.}\ \bibnamefont {Germanenko}},\ }\bibfield
  {title} {\bibinfo {title} {Transformation of energy spectrum and wave
  functions on the way from a 2d-to-3d topological insulator in hgte quantum
  wells},\ }\href {https://doi.org/10.1103/PhysRevB.106.085301} {\bibfield
  {journal} {\bibinfo  {journal} {Phys. Rev. B}\ }\textbf {\bibinfo {volume}
  {106}},\ \bibinfo {pages} {085301} (\bibinfo {year} {2022})}\BibitemShut
  {NoStop}%
\bibitem [{\citenamefont {Simon}\ and\ \citenamefont
  {Halperin}(1994)}]{Simon94}%
  \BibitemOpen
  \bibfield  {author} {\bibinfo {author} {\bibfnamefont {S.~H.}\ \bibnamefont
  {Simon}}\ and\ \bibinfo {author} {\bibfnamefont {B.~I.}\ \bibnamefont
  {Halperin}},\ }\bibfield  {title} {\bibinfo {title} {Explanation for the
  resistivity law in quantum hall systems},\ }\href
  {https://doi.org/10.1103/PhysRevLett.73.3278} {\bibfield  {journal} {\bibinfo
   {journal} {Phys. Rev. Lett.}\ }\textbf {\bibinfo {volume} {73}},\ \bibinfo
  {pages} {3278} (\bibinfo {year} {1994})}\BibitemShut {NoStop}%
\bibitem [{\citenamefont {Mani}\ \emph {et~al.}(2009)\citenamefont {Mani},
  \citenamefont {Johnson}, \citenamefont {Umansky}, \citenamefont
  {Narayanamurti},\ and\ \citenamefont {Ploog}}]{Mani09}%
  \BibitemOpen
  \bibfield  {author} {\bibinfo {author} {\bibfnamefont {R.~G.}\ \bibnamefont
  {Mani}}, \bibinfo {author} {\bibfnamefont {W.~B.}\ \bibnamefont {Johnson}},
  \bibinfo {author} {\bibfnamefont {V.}~\bibnamefont {Umansky}}, \bibinfo
  {author} {\bibfnamefont {V.}~\bibnamefont {Narayanamurti}},\ and\ \bibinfo
  {author} {\bibfnamefont {K.}~\bibnamefont {Ploog}},\ }\bibfield  {title}
  {\bibinfo {title} {Phase study of oscillatory resistances in
  microwave-irradiated- and dark-gaas/algaas devices: Indications of an
  unfamiliar class of the integral quantum hall effect},\ }\href
  {https://doi.org/10.1103/PhysRevB.79.205320} {\bibfield  {journal} {\bibinfo
  {journal} {Phys. Rev. B}\ }\textbf {\bibinfo {volume} {79}},\ \bibinfo
  {pages} {205320} (\bibinfo {year} {2009})}\BibitemShut {NoStop}%
\bibitem [{\citenamefont {Qian}\ \emph {et~al.}(2017)\citenamefont {Qian},
  \citenamefont {Nakamura}, \citenamefont {Fallahi}, \citenamefont {Gardner},
  \citenamefont {Watson}, \citenamefont {L\"uscher}, \citenamefont {Folk},
  \citenamefont {Cs\'athy},\ and\ \citenamefont {Manfra}}]{Qian17}%
  \BibitemOpen
  \bibfield  {author} {\bibinfo {author} {\bibfnamefont {Q.}~\bibnamefont
  {Qian}}, \bibinfo {author} {\bibfnamefont {J.}~\bibnamefont {Nakamura}},
  \bibinfo {author} {\bibfnamefont {S.}~\bibnamefont {Fallahi}}, \bibinfo
  {author} {\bibfnamefont {G.~C.}\ \bibnamefont {Gardner}}, \bibinfo {author}
  {\bibfnamefont {J.~D.}\ \bibnamefont {Watson}}, \bibinfo {author}
  {\bibfnamefont {S.}~\bibnamefont {L\"uscher}}, \bibinfo {author}
  {\bibfnamefont {J.~A.}\ \bibnamefont {Folk}}, \bibinfo {author}
  {\bibfnamefont {G.~A.}\ \bibnamefont {Cs\'athy}},\ and\ \bibinfo {author}
  {\bibfnamefont {M.~J.}\ \bibnamefont {Manfra}},\ }\bibfield  {title}
  {\bibinfo {title} {Quantum lifetime in ultrahigh quality gaas quantum wells:
  Relationship to ${\mathrm{\ensuremath{\Delta}}}_{5/2}$ and impact of density
  fluctuations},\ }\href {https://doi.org/10.1103/PhysRevB.96.035309}
  {\bibfield  {journal} {\bibinfo  {journal} {Phys. Rev. B}\ }\textbf {\bibinfo
  {volume} {96}},\ \bibinfo {pages} {035309} (\bibinfo {year}
  {2017})}\BibitemShut {NoStop}%
\bibitem [{Note4()}]{Note4}%
  \BibitemOpen
  \bibinfo {note} {The inverted character of the spectrum in this structure is
  justified in our article \cite {Minkov16}, in which the intersection of
  Landau levels belonging to the conduction and valence bands, specific for
  quantum wells with an inverted spectrum, was observed.}\BibitemShut {Stop}%
\bibitem [{\citenamefont {Minkov}\ \emph {et~al.}(2005)\citenamefont {Minkov},
  \citenamefont {Sherstobitov}, \citenamefont {Germanenko}, \citenamefont
  {Rut}, \citenamefont {Larionova},\ and\ \citenamefont {Zvonkov.}}]{Min05}%
  \BibitemOpen
  \bibfield  {author} {\bibinfo {author} {\bibfnamefont {G.~M.}\ \bibnamefont
  {Minkov}}, \bibinfo {author} {\bibfnamefont {A.~A.}\ \bibnamefont
  {Sherstobitov}}, \bibinfo {author} {\bibfnamefont {A.~V.}\ \bibnamefont
  {Germanenko}}, \bibinfo {author} {\bibfnamefont {O.~E.}\ \bibnamefont {Rut}},
  \bibinfo {author} {\bibfnamefont {V.~A.}\ \bibnamefont {Larionova}},\ and\
  \bibinfo {author} {\bibfnamefont {B.~N.}\ \bibnamefont {Zvonkov.}},\
  }\bibfield  {title} {\bibinfo {title} {Antilocalization and spin-orbit
  coupling in hole strained gaas/in$_x$ga$_{1-x}$as/gaas quantum well
  heterostructures},\ }\href@noop {} {\bibfield  {journal} {\bibinfo  {journal}
  {Phys. Rev. B}\ }\textbf {\bibinfo {volume} {71}},\ \bibinfo {pages} {165312
  (1 } (\bibinfo {year} {2005})}\BibitemShut {NoStop}%
\bibitem [{\citenamefont {Dyakonov}\ and\ \citenamefont
  {Khaetskii}(2008)}]{Dyakonov08}%
  \BibitemOpen
  \bibfield  {author} {\bibinfo {author} {\bibfnamefont {M.~I.}\ \bibnamefont
  {Dyakonov}}\ and\ \bibinfo {author} {\bibfnamefont {A.~V.}\ \bibnamefont
  {Khaetskii}},\ }\bibinfo {title} {Spin hall effect},\ in\ \href
  {https://doi.org/10.1007/978-3-540-78820-1_8} {\emph {\bibinfo {booktitle}
  {Spin Physics in Semiconductors}}},\ \bibinfo {editor} {edited by\ \bibinfo
  {editor} {\bibfnamefont {M.~I.}\ \bibnamefont {Dyakonov}}}\ (\bibinfo
  {publisher} {Springer Berlin Heidelberg},\ \bibinfo {address} {Berlin,
  Heidelberg},\ \bibinfo {year} {2008})\ pp.\ \bibinfo {pages}
  {211--243}\BibitemShut {NoStop}%
\bibitem [{\citenamefont {Sinova}\ \emph {et~al.}(2015)\citenamefont {Sinova},
  \citenamefont {Valenzuela}, \citenamefont {Wunderlich}, \citenamefont
  {Back},\ and\ \citenamefont {Jungwirth}}]{Sinova15}%
  \BibitemOpen
  \bibfield  {author} {\bibinfo {author} {\bibfnamefont {J.}~\bibnamefont
  {Sinova}}, \bibinfo {author} {\bibfnamefont {S.~O.}\ \bibnamefont
  {Valenzuela}}, \bibinfo {author} {\bibfnamefont {J.}~\bibnamefont
  {Wunderlich}}, \bibinfo {author} {\bibfnamefont {C.~H.}\ \bibnamefont
  {Back}},\ and\ \bibinfo {author} {\bibfnamefont {T.}~\bibnamefont
  {Jungwirth}},\ }\bibfield  {title} {\bibinfo {title} {Spin hall effects},\
  }\href {https://doi.org/10.1103/RevModPhys.87.1213} {\bibfield  {journal}
  {\bibinfo  {journal} {Rev. Mod. Phys.}\ }\textbf {\bibinfo {volume} {87}},\
  \bibinfo {pages} {1213} (\bibinfo {year} {2015})}\BibitemShut {NoStop}%
\bibitem [{\citenamefont {Kuntsevich}\ \emph {et~al.}(2018)\citenamefont
  {Kuntsevich}, \citenamefont {Shupletsov},\ and\ \citenamefont
  {Minkov}}]{Kuntsevich18}%
  \BibitemOpen
  \bibfield  {author} {\bibinfo {author} {\bibfnamefont {A.~Y.}\ \bibnamefont
  {Kuntsevich}}, \bibinfo {author} {\bibfnamefont {A.~V.}\ \bibnamefont
  {Shupletsov}},\ and\ \bibinfo {author} {\bibfnamefont {G.~M.}\ \bibnamefont
  {Minkov}},\ }\bibfield  {title} {\bibinfo {title} {Simple mechanisms that
  impede the berry phase identification from magneto-oscillations},\ }\href
  {https://doi.org/10.1103/PhysRevB.97.195431} {\bibfield  {journal} {\bibinfo
  {journal} {Phys. Rev. B}\ }\textbf {\bibinfo {volume} {97}},\ \bibinfo
  {pages} {195431} (\bibinfo {year} {2018})}\BibitemShut {NoStop}%
\bibitem [{\citenamefont {Gerchikov}\ and\ \citenamefont
  {Subashiev}(1990)}]{Gerchikov90}%
  \BibitemOpen
  \bibfield  {author} {\bibinfo {author} {\bibfnamefont {L.~G.}\ \bibnamefont
  {Gerchikov}}\ and\ \bibinfo {author} {\bibfnamefont {A.}~\bibnamefont
  {Subashiev}},\ }\bibfield  {title} {\bibinfo {title} {Interface states in
  subbabnd structure of semiconductor quantum well},\ }\href@noop {} {\bibfield
   {journal} {\bibinfo  {journal} {Phys. Stat. Sol. (b)}\ }\textbf {\bibinfo
  {volume} {160}},\ \bibinfo {pages} {443} (\bibinfo {year}
  {1990})}\BibitemShut {NoStop}%
\bibitem [{\citenamefont {Zhang}\ \emph {et~al.}(2001)\citenamefont {Zhang},
  \citenamefont {Pfeuffer-Jeschke}, \citenamefont {Ortner}, \citenamefont
  {Hock}, \citenamefont {Buhmann}, \citenamefont {Becker},\ and\ \citenamefont
  {Landwehr}}]{Zhang01}%
  \BibitemOpen
  \bibfield  {author} {\bibinfo {author} {\bibfnamefont {X.~C.}\ \bibnamefont
  {Zhang}}, \bibinfo {author} {\bibfnamefont {A.}~\bibnamefont
  {Pfeuffer-Jeschke}}, \bibinfo {author} {\bibfnamefont {K.}~\bibnamefont
  {Ortner}}, \bibinfo {author} {\bibfnamefont {V.}~\bibnamefont {Hock}},
  \bibinfo {author} {\bibfnamefont {H.}~\bibnamefont {Buhmann}}, \bibinfo
  {author} {\bibfnamefont {C.~R.}\ \bibnamefont {Becker}},\ and\ \bibinfo
  {author} {\bibfnamefont {G.}~\bibnamefont {Landwehr}},\ }\bibfield  {title}
  {\bibinfo {title} {Rashba splitting in \textit{n} -type modulation-doped hgte
  quantum wells with an inverted band structure},\ }\href
  {https://doi.org/10.1103/PhysRevB.63.245305} {\bibfield  {journal} {\bibinfo
  {journal} {Phys. Rev. B}\ }\textbf {\bibinfo {volume} {63}},\ \bibinfo
  {pages} {245305} (\bibinfo {year} {2001})}\BibitemShut {NoStop}%
\bibitem [{\citenamefont {Novik}\ \emph {et~al.}(2005)\citenamefont {Novik},
  \citenamefont {Pfeuffer-Jeschke}, \citenamefont {Jungwirth}, \citenamefont
  {Latussek}, \citenamefont {Becker}, \citenamefont {Landwehr}, \citenamefont
  {Buhmann},\ and\ \citenamefont {Molenkamp}}]{Novik05}%
  \BibitemOpen
  \bibfield  {author} {\bibinfo {author} {\bibfnamefont {E.~G.}\ \bibnamefont
  {Novik}}, \bibinfo {author} {\bibfnamefont {A.}~\bibnamefont
  {Pfeuffer-Jeschke}}, \bibinfo {author} {\bibfnamefont {T.}~\bibnamefont
  {Jungwirth}}, \bibinfo {author} {\bibfnamefont {V.}~\bibnamefont {Latussek}},
  \bibinfo {author} {\bibfnamefont {C.~R.}\ \bibnamefont {Becker}}, \bibinfo
  {author} {\bibfnamefont {G.}~\bibnamefont {Landwehr}}, \bibinfo {author}
  {\bibfnamefont {H.}~\bibnamefont {Buhmann}},\ and\ \bibinfo {author}
  {\bibfnamefont {L.~W.}\ \bibnamefont {Molenkamp}},\ }\bibfield  {title}
  {\bibinfo {title} {Band structure of semimagnetic hg$_{1-y}$mn$_y$te quantum
  wells},\ }\href {https://doi.org/10.1103/PhysRevB.72.035321} {\bibfield
  {journal} {\bibinfo  {journal} {Phys. Rev. B}\ }\textbf {\bibinfo {volume}
  {72}},\ \bibinfo {pages} {035321} (\bibinfo {year} {2005})}\BibitemShut
  {NoStop}%
\bibitem [{\citenamefont {Bernevig}\ \emph {et~al.}(2006)\citenamefont
  {Bernevig}, \citenamefont {Hughes},\ and\ \citenamefont
  {Zhang}}]{Bernevig06}%
  \BibitemOpen
  \bibfield  {author} {\bibinfo {author} {\bibfnamefont {B.~A.}\ \bibnamefont
  {Bernevig}}, \bibinfo {author} {\bibfnamefont {T.~L.}\ \bibnamefont
  {Hughes}},\ and\ \bibinfo {author} {\bibfnamefont {S.-C.}\ \bibnamefont
  {Zhang}},\ }\bibfield  {title} {\bibinfo {title} {Quantum spin hall effect
  and topological phase transition in hgte quantum wells},\ }\href
  {https://doi.org/10.1126/science.1133734} {\bibfield  {journal} {\bibinfo
  {journal} {Science}\ }\textbf {\bibinfo {volume} {314}},\ \bibinfo {pages}
  {1757} (\bibinfo {year} {2006})}\BibitemShut {NoStop}%
\bibitem [{\citenamefont {Zholudev}(2013)}]{ZholudevPhD}%
  \BibitemOpen
  \bibfield  {author} {\bibinfo {author} {\bibfnamefont {M.}~\bibnamefont
  {Zholudev}},\ }\href@noop {} {Ph.D. thesis},\ \bibinfo  {school} {University
  Montpellier 2, France} (\bibinfo {year} {2013})\BibitemShut {NoStop}%
\bibitem [{\citenamefont {Ren}\ \emph {et~al.}(2016)\citenamefont {Ren},
  \citenamefont {Qiao},\ and\ \citenamefont {Niu}}]{Ren2016}%
  \BibitemOpen
  \bibfield  {author} {\bibinfo {author} {\bibfnamefont {Y.}~\bibnamefont
  {Ren}}, \bibinfo {author} {\bibfnamefont {Z.}~\bibnamefont {Qiao}},\ and\
  \bibinfo {author} {\bibfnamefont {Q.}~\bibnamefont {Niu}},\ }\bibfield
  {title} {\bibinfo {title} {Topological phases in two-dimensional materials: a
  review},\ }\href {http://stacks.iop.org/0034-4885/79/i=6/a=066501} {\bibfield
   {journal} {\bibinfo  {journal} {Reports on Progress in Physics}\ }\textbf
  {\bibinfo {volume} {79}},\ \bibinfo {pages} {066501} (\bibinfo {year}
  {2016})}\BibitemShut {NoStop}%
\bibitem [{\citenamefont {D'yakonov}\ and\ \citenamefont
  {Khaetskii}(1982)}]{Dyak82e}%
  \BibitemOpen
  \bibfield  {author} {\bibinfo {author} {\bibfnamefont {M.~I.}\ \bibnamefont
  {D'yakonov}}\ and\ \bibinfo {author} {\bibfnamefont {A.}~\bibnamefont
  {Khaetskii}},\ }\href@noop {} {\bibfield  {journal} {\bibinfo  {journal} {Zh.
  Eksp. Teor. Fiz.}\ }\textbf {\bibinfo {volume} {82}},\ \bibinfo {pages}
  {1584} (\bibinfo {year} {1982})},\ \translation{Sov. Phys. JETP \textbf{55},
  917 (1982)}\BibitemShut {NoStop}%
\bibitem [{\citenamefont {Minkov}\ \emph {et~al.}(2016)\citenamefont {Minkov},
  \citenamefont {Germanenko}, \citenamefont {Rut}, \citenamefont
  {Sherstobitov}, \citenamefont {Nestoklon}, \citenamefont {Dvoretski},\ and\
  \citenamefont {Mikhailov}}]{Minkov16}%
  \BibitemOpen
  \bibfield  {author} {\bibinfo {author} {\bibfnamefont {G.~M.}\ \bibnamefont
  {Minkov}}, \bibinfo {author} {\bibfnamefont {A.~V.}\ \bibnamefont
  {Germanenko}}, \bibinfo {author} {\bibfnamefont {O.~E.}\ \bibnamefont {Rut}},
  \bibinfo {author} {\bibfnamefont {A.~A.}\ \bibnamefont {Sherstobitov}},
  \bibinfo {author} {\bibfnamefont {M.~O.}\ \bibnamefont {Nestoklon}}, \bibinfo
  {author} {\bibfnamefont {S.~A.}\ \bibnamefont {Dvoretski}},\ and\ \bibinfo
  {author} {\bibfnamefont {N.~N.}\ \bibnamefont {Mikhailov}},\ }\bibfield
  {title} {\bibinfo {title} {Spin-orbit splitting of valence and conduction
  bands in hgte quantum wells near the dirac point},\ }\href
  {https://doi.org/10.1103/PhysRevB.93.155304} {\bibfield  {journal} {\bibinfo
  {journal} {Phys. Rev. B}\ }\textbf {\bibinfo {volume} {93}},\ \bibinfo
  {pages} {155304} (\bibinfo {year} {2016})}\BibitemShut {NoStop}%
\end{thebibliography}
%
\end{document}